\documentclass[12pt,a4paper]{article}
\usepackage{a4wide}
\usepackage{epsfig}
\usepackage{amsmath}
\usepackage{latexsym}
\usepackage{latexsym}
\usepackage{array}
  \newlength{\absize}
\newcommand{\dd}{\mbox{{\rm d}}}

\def\lsim{\mathrel{\rlap{\raise 2.5pt \hbox{$<$}}\lower 2.5pt
\hbox{$\sim$}}}

\newcommand{\Lumint}{{\cal L}_{\rm int}}

\allowdisplaybreaks 

\catcode`@=11
\def\citer{\@ifnextchar [{\@tempswatrue\@citexr}{\@tempswafalse\@citexr[]}}

\def\@citexr[#1]#2{\if@filesw\immediate\write\@auxout{\string\citation{#2}}\fi
  \def\@citea{}\@cite{\@for\@citeb:=#2\do
    {\@citea\def\@citea{--\penalty\@m}\@ifundefined
       {b@\@citeb}{{\bf ?}\@warning
       {Citation `\@citeb' on page \thepage \space undefined}}%
\hbox{\csname b@\@citeb\endcsname}}}{#1}} \catcode`@=12

\begin{document}
  \thispagestyle{empty}
  \pagestyle{empty}
  \renewcommand{\thefootnote}{\fnsymbol{footnote}}
\newpage\normalsize
    \pagestyle{plain}
    \setlength{\baselineskip}{4ex}\par
    \setcounter{footnote}{0}
    \renewcommand{\thefootnote}{\arabic{footnote}}
\newcommand{\preprint}[1]{%
  \begin{flushright}
    \setlength{\baselineskip}{3ex} #1
  \end{flushright}}
\renewcommand{\title}[1]{%
  \begin{center}
    \LARGE #1
  \end{center}\par}
\renewcommand{\author}[1]{%
  \vspace{2ex}
  {\Large
   \begin{center}
     \setlength{\baselineskip}{3ex} #1 \par
   \end{center}}}
\renewcommand{\thanks}[1]{\footnote{#1}}
\renewcommand{\abstract}[1]{%
  \vspace{2ex}
  \normalsize
  \begin{center}
    \centerline{\bf Abstract}\par
    \vspace{2ex}
    \parbox{\absize}{#1\setlength{\baselineskip}{2.5ex}\par}
  \end{center}}

\begin{flushright}
{\setlength{\baselineskip}{2ex}\par

}
\end{flushright}
\vspace*{4mm}
\vfill
\title{Distinguishing New Physics Scenarios at a Linear Collider
with Polarized Beams}

\vfill

\author{
A.A. Pankov$^{a}$, N. Paver$^{b}$ {\rm and}  A.V.
Tsytrinov$^{a}$ }
\begin{center}
$^a$ ICTP Affiliated Centre, Pavel Sukhoi Technical University,
     Gomel 246746, Belarus \\
$^b$ University of Trieste  and INFN-Sezione di Trieste, 34100
Trieste, Italy \\
\end{center}
\vfill \abstract {Numerous non-standard dynamics are described by
contact-like effective interactions that can manifest themselves
only through deviations of the cross sections from the Standard
Model predictions. If one such deviation were observed, it should
be important to definitely identify, to a given confidence level,
the actual source among the possible non-standard interactions
that in principle can explain it. We here estimate the
``identification'' reach on different New Physics effective
interactions obtainable from angular distributions of lepton pair
production processes at the planned International Linear Collider
with polarized beams. The models for which we discuss the range in
the relevant high mass scales where they can be ``identified'' as
sources of corrections from the Standard Model predictions, are the
interactions based on gravity in large and in ${\rm TeV}^{-1}$
extra dimensions and the compositeness-inspired four-fermion contact
interactions. The availability of both beams polarized in many cases
plays an essential r{\^o}le in enhancing the identification sensitivity.}

\vspace*{20mm}
\setcounter{footnote}{0}
\vfill

\newpage
    \setcounter{footnote}{0}
    \renewcommand{\thefootnote}{\arabic{footnote}}

\section{Introduction}\label{sec:I}
Numerous new physics (NP) scenarios, candidates as solutions of
Standard Model (SM) conceptual problems, are characterized by
novel interactions mediated by exchanges of very heavy states with
mass scales significantly greater than the electroweak scale. In
many cases, theoretical considerations as well as current
experimental constraints indicate that the new objects may be too
heavy to be directly produced even at the highest energies
foreseen at future colliders, such as the CERN Large Hadron
Collider (LHC) and the $e^+-e^-$ International Linear Collider
(ILC). In this situation the new, non-standard, interactions could
only be revealed by indirect, virtual, effects manifesting
themselves as deviations of measured cross sections from the
SM predictions.
\par
At the available ``low'' energies provided by the accelerators,
where we study reactions among the familiar SM particles,
effective ``contact'' interaction Lagrangians represent the
convenient theoretical tool to physically parameterize the effects
of the above mentioned non-standard interactions and, in particular,
to test the corresponding virtual high mass exchanges. Clearly, in
this framework the transition amplitudes are parameterized as
power expansions in the (small) ratios between the Mandelstam
variables of the process under study and the high mass scales
squared. The sensitivity to the searched for signals will
therefore be increased by the colliders high energy (and high
luminosity).
\par
Since, in principle, the observation of a correction to the SM
cross section may by itself not enable to unambiguously identify
its source among the different possible explanations, suitable
observables with enhanced sensitivity to the individual
non-standard scenarios and/or convenient statistical criteria in
the data analysis must be defined in order to discriminate the
nature of the relevant new physics model against the others. As
regards the search of effective contact interactions at the
planned high energy colliders, it should therefore be desirable to
assess for each non-standard scenario not only the ``discovery
reach'', i.e., the maximum value of the relevant mass scale below
which it produces observable corrections to the SM predictions,
but also the upper limit of the range of mass scale values where the
considered scenario not only produces observed deviations,
but can also be discriminated from the other potential sources of
the deviations themselves. We may call it the ``identification reach'' of the
considered model. Accordingly, it should be important to try to
achieve, for the various models, the maximum sensitivity (i.e.,
discovery reach) as well as highest possible identification reach.
\par
Here, we will try to quantitatively discuss the above issues in the cases of
the Kaluza-Klein (KK) graviton exchange in the context of gravity
propagating in ``large'' compactified extra
dimensions \citer{Arkani-Hamed:1998rs,Antoniadis:1998ig}, and of the
four-fermion contact interactions inspired by the context of leptons
and quark compositeness
\citer{'tHooft:xb,Ruckl:1983hz}.\footnote{Actually, this kind of description
more generally applies to a variety of new interactions generated by
exchanges of very massive virtual objects such as, for example, $Z^\prime$s,
leptoquarks and heavy scalars.} In particular, our aim will be to assess the
potential of electron and positron longitudinal polarization at the ILC, to
enhance the identification reaches on these NP effective interactions.
\par
Specifically, to this purpose we will take as basic observables the polarized
angular differential cross sections of the purely leptonic processes, namely,
the Bhabha scattering:
\begin{equation}
e^++e^-\to e^++e^-, \label{bhabha}
\end{equation}
the annihilation into lepton pairs ($l=\mu,\,\tau$):
\begin{equation}
e^++e^-\to l^++l^- \label{annil},
\end{equation}
and the M{\o}ller scattering:
\begin{equation}
e^-+e^-\to e^-+e^-. \label{moller}
\end{equation}
Processes (\ref{bhabha})--(\ref{moller}) can all receive contributions from
graviton exchange and from four-fermion contact interactions as well, and
represent sensitive probes of the above mentioned scenarios,
particularly in the case of polarized initial beams
\citer{Pankov:2005ar,Pankov:2002qk}. Indeed, the interesting feature of the
ILC \cite{Aguilar-Saavedra:2001rg} is that polarization of both initial
beams can be available \cite{Moortgat-Pick:2005cw}, and that in principle
this collider could be run in both required modes, nemely,
$e^+e^-$ and $e^-e^-$. In Ref.~\cite{Pasztor:2001hc}, the identification
reach on the different contact interactions was studied by using a
Monte Carlo-based analysis of the unpolarized differential cross sections
of processes (\ref{bhabha}) and (\ref{annil}). Electron and positron
longitudinal polarization can substantially help in reducing the
``confusion'' regions of the parameters where models cannot be
discriminated from each other on a $\chi^2$ basis. Accordingly, the
identification reach on the individual models should be enhanced.
\par
In Sec. 2 we give the relevant expressions for the differential
polarized angular distributions and for the deviations from the SM
predictions in the different NP scenarios considered heres. Secs. 3 and 4
outline the numerical analysis and present the numerical results for the
discovery reaches and the distinction among the models, respectively, as
allowed by initial beams polarization. Finally, some conclusive remarks are
given in Sec. 5.

\section{Angular distributions and deviations from the SM}\label{sec:II}
The polarized differential cross section of process (\ref{bhabha}) can be
conveniently written as (see for example
\citer{Schrempp:1987zy,Cullen:2000ef} and \cite{Pankov:2002qk}):
\begin{eqnarray}
\frac{\dd\sigma(P^-,P^+)}{\dd\cos\theta}
&=&\frac{(1+P^-)\,(1-P^+)}4\,\frac{\dd\sigma_{\rm
R}}{\dd\cos\theta}+ \frac{(1-P^-)\,(1+P^+)}4\,\frac{\dd\sigma_{\rm
L}}{\dd\cos\theta}
\nonumber \\
&+&\frac{(1+P^-)\,(1+P^+)}4\,\frac{\dd\sigma_{{\rm
RL},t}}{\dd\cos\theta}+ \frac{(1-P^-)\,(1-P^+)}4
\,\frac{\dd\sigma_{{\rm LR},t}}{\dd\cos\theta}, \label{cross}
\end{eqnarray}
with $\theta$ the angle between incoming and outgoing
electrons in the c.m.\ frame and $P^\mp$ the longitudinal polarization of
electron and positron beams, respectively. In Eq.~(\ref{cross}):
\begin{eqnarray}
\frac{\dd\sigma_{\rm L}}{\dd\cos\theta}&=& \frac{\dd\sigma_{{\rm
LL}}}{\dd\cos\theta}+ \frac{\dd\sigma_{{\rm
LR},s}}{\dd\cos\theta},
\nonumber \\
\frac{\dd\sigma_{\rm R}}{\dd\cos\theta}&=& \frac{\dd\sigma_{{\rm
RR}}}{\dd\cos\theta}+ \frac{\dd\sigma_{{\rm
RL},s}}{\dd\cos\theta}, \label{sigP}
\end{eqnarray}
where
\begin{eqnarray}
\frac{\dd\sigma_{{\rm LL}}}{\dd\cos\theta}&=&
\frac{2\pi\alpha_{\rm e.m.}^2}{s}\,\big\vert G_{{\rm LL},s}+
G_{{\rm LL},t} \big\vert^2, \quad \frac{\dd\sigma_{{\rm
RR}}}{\dd\cos\theta}= \frac{2\pi\alpha_{\rm e.m.}^2}{s}\,\big\vert
G_{{\rm RR},s}+G_{{\rm RR},t}\big\vert^2 ,
\nonumber \\
\frac{\dd\sigma_{{\rm
LR},t}}{\dd\cos\theta}&=&\frac{\dd\sigma_{{\rm
RL},t}}{\dd\cos\theta}= \frac{2\pi\alpha_{\rm
e.m.}^2}{s}\,\big\vert G_{{\rm LR},t}\big\vert^2,
\quad\frac{\dd\sigma_{{\rm LR},s}}{\dd\cos\theta}=
\frac{\dd\sigma_{{\rm RL},s}}{\dd\cos\theta}=
\frac{2\pi\alpha_{\rm e.m.}^2}{s}\,\big\vert G_{{\rm
LR},s}\big\vert^2. \label{helsig}
\end{eqnarray}
The helicity amplitudes $G_{\alpha\beta}$ ($\alpha,\beta={\rm L,R}$) can
be expressed in terms of the SM $\gamma$ and $Z$ exchanges in the $s$-
and $t$-channels, plus deviations due to contact interactions representing the
physics beyond the SM, as follows:
\begin{eqnarray}
G_{{\rm LL},s}&=&{u}\,\left(\frac{1}{s}+\frac{g_{\rm L}^2
}{s-M^2_Z}+ \Delta_{{\rm LL},s}\right), \quad G_{{\rm
LL},t}={u}\,\left(\frac{1}{t}+\frac{g_{\rm L}^2}{t-M^2_Z}+
\Delta_{{\rm LL},t}\right),
\nonumber \\
G_{{\rm RR},s}&=&{u}\,\left(\frac{1}{s}+ \frac{g_{\rm
R}^2}{s-M^2_Z}+ \Delta_{{\rm RR},s}\right), \quad G_{{\rm
RR},t}={u}\,\left(\frac{1}{t}+\frac{g_{\rm R}^2}{t-M^2_Z}+
\Delta_{{\rm RR},t}\right),
\nonumber \\
G_{{\rm LR},s}&=&{t}\,\left(\frac{1}{s}+\frac{g_{\rm R}\hskip 2pt
g_{\rm L}}{s-M^2_Z}+\Delta_{{\rm LR},s}\right), \qquad G_{{\rm
LR},t}= s\,\left(\frac{1}{t}+\frac{g_{\rm R}\hskip 2pt g_{\rm
L}}{t-M^2_Z}+\Delta_{{\rm LR},t}\right). \label{helamp}
\end{eqnarray}
Here: $u,t=-s(1\pm\cos\theta)/2$; $M_Z$ represents the mass of the
$Z$; $g_{\rm R}=\tan\theta_W$, $g_{\rm L}=-\cot{2\,\theta_W}$ are
the SM right- and left-handed electron couplings to the $Z$, with
$\theta_W$ the electroweak mixing angle.
\par
The advantage of Eqs.~(\ref{cross})--(\ref{helamp}) is that the
expression for the differential cross section of process
(\ref{moller}) can be obtained directly from crossing symmetry: one has to
replace $s\leftrightarrow u$, $(P^-,P^+)\rightarrow (P^-_1,-P^-_2)$ with
$P^-_1$ and $P^-_2$ denoting the polarizations of the two initial electrons,
and divide the cross section by 2 to account for identical particles.
Also, the same equations are easy to adapt to the cross section of the
annihilation process (\ref{annil}), one simply must drop all
$t$-channel poles (actually, for this process, in principle
$\epsilon_{\rm LR}\ne \epsilon_{\rm RL}$ \cite{Schrempp:1987zy}).
\par
Turning to the amplitudes deviations $\Delta_{\alpha\beta}$ in
Eq.~(\ref{helamp}), in the ADD graviton exchange scenario
\citer{Arkani-Hamed:1998rs,Antoniadis:1998ig} only gravity can propagate in
two or more extra spatial dimensions of the millimeter order, whereas the SM
particles must live only in the ordinary four-dimensional
space.\footnote{Actually, while the case of two extra dimensions may be
marginal due to cosmological arguments and direct gravity experiments, three
extra dimensions have been advocated from dark-matter
observations \cite{Qin:2005pf}.} Massless graviton exchange in extra
dimensions translates to the exchange of a tower of evenly spaced KK massive
states (with vertices given in \cite{Giudice:1998ck,Han:1998sg}, see also
\cite{Giudice:2004mg}), and this effect can be parameterized by the effective,
dimension-8, contact interaction Lagrangian
\cite{Hewett:1998sn}
\begin{equation}
{\cal L}^{\rm
ADD}=i\frac{4\lambda}{\Lambda_H^4}T^{\mu\nu}T_{\mu\nu}.
\label{dim-8}
\end{equation}
Here, $T_{\mu\nu}$ is the energy-momentum tensor of the SM particles,
$\Lambda_H$ is a ultraviolet cut-off on the summation over the KK spectrum,
expected in the (multi) TeV region, and $\lambda=\pm1$ (models with
$\lambda=1$ or $\lambda=-1$ are denoted as ADD+ and ADD$-$, respectively). The
explicit expressions of the corresponding corrections to the SM
amplitudes relevant to Bhabha scattering are reported in
Table~\ref{table:epsilon}. In this table, we
include also the deviations corresponding to models with
${\rm TeV}^{-1}$-scale extra dimensions, parameterized by the
``compactification scale'' $M_C$, where also the SM gauge bosons may
propagate in the additional dimensions \cite{Cheung:2001mq,Rizzo:1999br}.
\begin{table}[htb]
\centering \caption{Parametrization of the $\Delta_{\alpha\beta}$
functions in different models for $e^+e^-\to e^+e^-$}
\vspace{0.3cm}
\begin{tabular}{|l||c|}
\hline new physics model & $\Delta_{\alpha\beta}$  \\  \hline
\hline composite fermions \cite{Eichten:1983hw} &
${\displaystyle{\Delta_{{\alpha\beta},s}=\Delta_{{\alpha\beta},t}=
\frac{1}{\alpha_{\rm e.m.}}\frac{\eta_{\alpha\beta}} {\Lambda_
{\alpha\beta}^2}}}$
\\
\hline
${\rm TeV}^{-1}$-scale extra dim. \cite{Cheung:2001mq,Rizzo:1999br} &
${\displaystyle{\Delta_{{\alpha\beta},s}=\Delta_{{\alpha\beta},t}=
-(Q_eQ_f+g_\alpha^e\,g_\beta^f)\,\frac{\pi^2}{3\,M_C^2}}}$
 \\
 \hline
 & ${\displaystyle{\Delta_{{\rm LL},s}=\Delta_{{\rm RR},s}=
\frac{\lambda}{\pi\alpha_{\rm e.m.}\Lambda_H^4} (u+\frac{3}{4}s)}}$ \\
 & ${\displaystyle{\Delta_{{\rm LL},t}=\Delta_{{\rm RR},t}=
\frac{\lambda}{\pi\alpha_{\rm e.m.}\Lambda_H^4} (u+\frac{3}{4}t)}}$ \\
ADD model \cite{Pasztor:2001hc,Cullen:2000ef} &
${\displaystyle{\Delta_{{\rm LR},s}=-\frac{\lambda}{\pi\alpha_{\rm
e.m.}\Lambda_H^4} (t+\frac{3}{4}s)}}$ \\
&   ${\displaystyle{\Delta_{{\rm
LR},t}=-\frac{\lambda}{\pi\alpha_{\rm e.m.}\Lambda_H^4}
(s+\frac{3}{4}t)}}$
\\  \hline
\end{tabular}
\label{table:epsilon}
\end{table}
Current experimental limits, from LEP2 and Tevatron, are in the range
$\Lambda_H>1.1-1.3$ TeV \citer{Ask:2004dv,Abazov:2005tk}. For the
${\rm TeV}^{-1}$-scale extra dimension scenario the limit, mostly determined
by LEP data, is $M_C>6.8\hskip 3pt{\rm TeV}$ \cite{Cheung:2004ab}.
\par
The four-fermion contact interaction scenario can be represented by the
following vector-vector, dimension-6, effective Lagrangian (with
$\eta_{\alpha\beta}=\pm 1,0$ and $\alpha,\beta={\rm L,R}$)
\cite{Eichten:1983hw}:
\begin{equation}
{\cal L}^{\rm CI}=\frac{4\pi}{1+\delta_{ef}}\hskip 3pt \sum_{\alpha,\beta}
\hskip 3pt
\frac{\eta_{\alpha\beta}}{\Lambda^2_{\alpha\beta}} \left(\bar
e_\alpha\gamma_\mu e_\alpha\right) \left(\bar f_\beta\gamma^\mu
f_\beta\right), \label{CI}
\end{equation}
where $\Lambda_{\alpha\beta}$ denote compositeness scales and
$\delta_{ef}=$1 (0) for $f=e$ ($f\ne e$). The most popular four-fermion
contact interaction models (CI) are defined by specializing, in
Eq.~(\ref{CI}), the helicities according to Table~\ref{table:composit}.
\begin{table}[htb]
\label{table:composit}
\caption{Definition of four-fermion CI models}
\begin{center}
\begin{tabular}{|c||c|c|c|c|} \hline
CI model & $\eta_{\rm LL}$ & $\eta_{\rm RR}$ & $\eta_{\rm LR}$ &
$\eta_{\rm RL}$  \\
\hline \hline LL & $\pm 1$ & 0 & 0 & 0 \\ \hline RR & 0 & $\pm 1$
  &  0 & 0 \\ \hline LR & 0 & 0 & $\pm 1$ & 0 \\ \hline RL & 0 & 0 &
0 & $\pm 1$ \\ \hline VV & $\pm 1$ & $\pm 1$ & $\pm 1$ & $\pm 1$
\\ \hline
AA & $\pm 1$ & $\pm 1$ & $\mp 1$ & $\mp 1$ \\ \hline
\end{tabular}
\end{center}
\end{table}
\par\noindent
The corresponding deviations $\Delta_{\alpha\beta}$, that appear
in Eq.~(\ref{helamp}), are listed in Table~\ref{table:epsilon}.
Current limits on $\Lambda$s significantly vary according the
process studied and the kind of analysis performed there. In
general, the lower bounds are of the order of $10\hskip 2pt{\rm
TeV}$ (a detailed presentation can be found in the listings of
Ref.~\cite{Eidelman}).
\par
It should be noted from Table~\ref{table:epsilon} that, contrary
to the other cases, for the ADD scenario the amplitudes deviations
are $z$-dependent ($z\equiv\cos\theta$), and consequently add
extra terms, proportional to $z^3$ and $z^4$, to the SM angular
distribution of the annihilation process (\ref{annil}). However,
these contributions to the total cross section are expected to be
tiny, because the (expected dominant) interference with the SM
amplitudes vanishes when integrated over the full angular range
and there remain only the terms quadratic in (\ref{dim-8}),
suppressed by the corresponding very high power of the small
parameter $1/\Lambda_H$. On the contrary, the interference between
graviton exchange and $t$-channel SM exchanges for process
(\ref{bhabha}) may give non-vanishing contributions, which may
favorably combine with the larger statistical precisions expected
in this channel and increase the discovery reach on $\Lambda_H$.

\section{Discovery reach on the contact interaction models}\label{sec:III}
We here briefly outline the derivation of the expected discovery reaches
on the New Physics scenarios introduced in the previous section. The basic
objects are the relative deviations of observables from the SM predictions
due to the NP:
\begin{equation}
\Delta ({\cal O})=\frac{{\cal O}(\rm SM+NP)-{\cal O}(\rm SM)}{{\cal
O}(\rm SM)},
\label{relat}
\end{equation}
and, as anticipated, we concentrate on the differential cross section,
${\cal O}\equiv\dd\sigma/\dd\cos\theta$. To get an illustration of the effects
induced by the individual NP models, we show in
Figs.~1--3 the angular behavior of the relative deviations
(\ref{relat}) for the three leptonic processes under consideration
(with unpolarized beams), for c.m. energy $\sqrt s=0.5$ TeV and
selected values of the relevant mass scale parameters close to
their discovery reaches (unpolarized cross sections). The
superscript ``+'' on the CI mass scales $\Lambda_{\alpha\beta}$
denotes the choice $\eta_{\alpha\beta}=1$ in
Table~\ref{table:composit}, while the notation ADD$\pm$
corresponds to $\lambda=\pm 1$ in Eq.~(\ref{dim-8}). Vertical bars
represent the statistical uncertainty in each angular bin, for an
integrated luminosity ${\cal L}_{\rm int}=100\, {\rm fb}^{-1}$. Of
course, the comparison of deviations with statistical
uncertainties is an indicator of the sensitivity of an observable
to the individual effective interaction models.
\begin{figure}[htb!]
\label{bhab}
\begin{center}
\epsfig{file=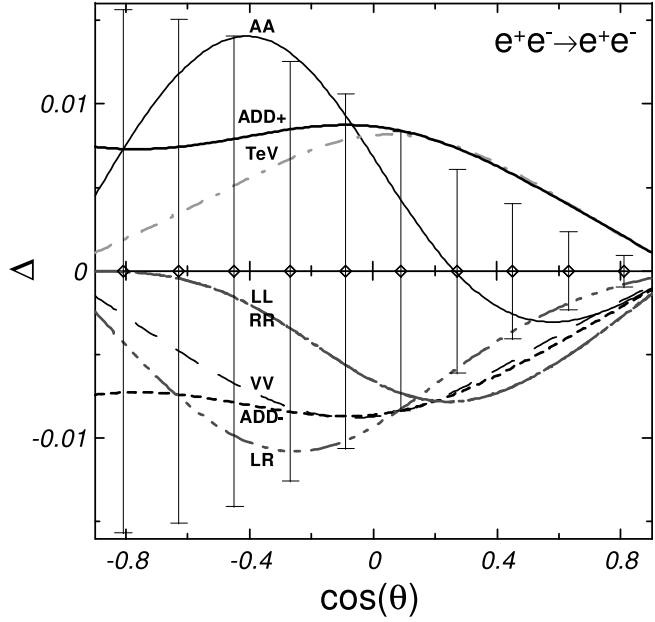,width=9.cm} \vspace{-2.5cm}
\caption{Relative deviations of the unpolarized Bhabha differential
cross section from the SM prediction as a function of $\cos\theta$ at
$\sqrt s=0.5$ TeV for the CI models of Table~\ref{table:composit}:
AA ($\Lambda_{\rm AA}^+$=48 TeV), VV ($\Lambda_{\rm VV }^+$=76 TeV),
LL ($\Lambda_{\rm LL}^+$=37 TeV), RR ($\Lambda_{\rm RR}^+$=36 TeV),
LR ($\Lambda_{\rm LR}^+$=60 TeV); for the ${\rm TeV}^{-1}$ model
($M_C$=12 TeV) and the ADD$\pm$ models ($\Lambda_H$=4 TeV). The vertical
bars represent the statistical uncertainty in each bin for
${\cal L}_{\rm int}=100\, {\rm fb}^{-1}$.}
\end{center}
\end{figure}
\begin{figure}[htb!]
\label{mm}
\begin{center}
\vspace{-2.cm}
\epsfig{file=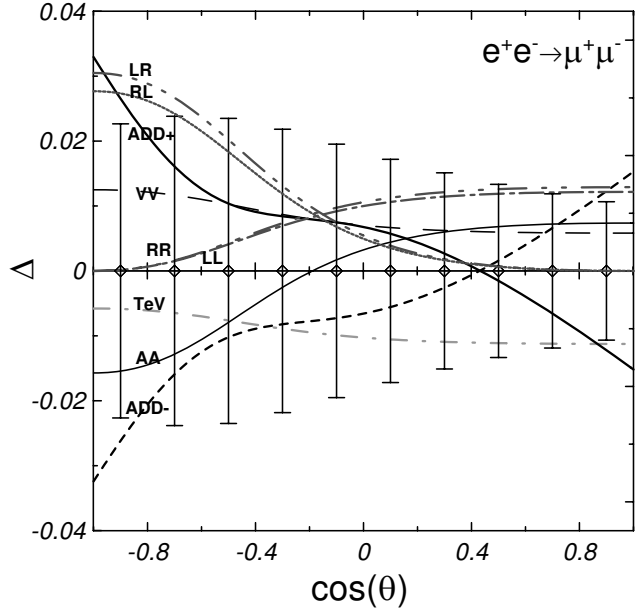,width=9.cm} \vspace{-2.5cm} \caption{
Same as in Fig.~1 but for $e^+e^-\to \mu^+\mu^-$, for the CI
models of Table~\ref{table:composit}: AA ($\Lambda_{\rm AA}^+$=80
TeV), VV ($\Lambda_{\rm VV }^+$=90 TeV), LL ($\Lambda_{\rm
LL}^+$=45 TeV), RR ($\Lambda_{\rm RR}^+$=42 TeV), LR
($\Lambda_{\rm LR}^+$=41 TeV), RL ($\Lambda_{\rm RL}^+$=43 TeV);
for the ${\rm TeV}^{-1}$ model ($M_C$=17 TeV) and the ADD$\pm$
models ($\Lambda_H$=2.8 TeV).}
\end{center}
\end{figure}
\begin{figure}[htb!]
\label{mol}
\begin{center}
\vspace{-2.cm}
\epsfig{file=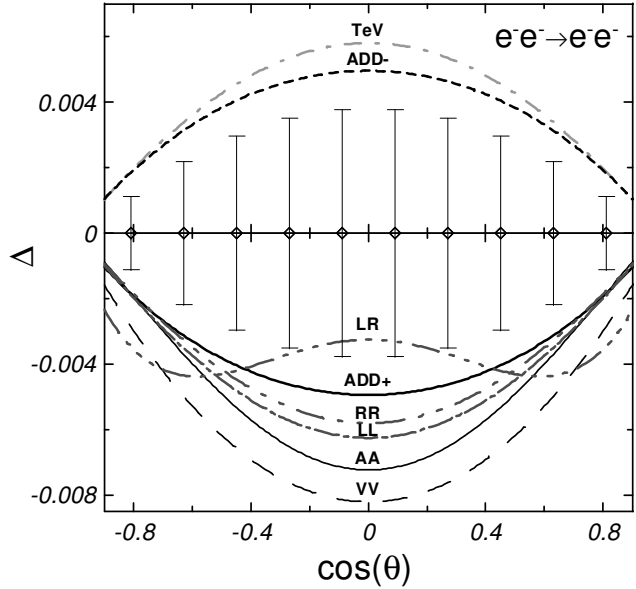,width=9.cm} \vspace{-2.5cm} \caption{Same
as in Fig.~1 but for M{\o}ller scattering for the CI models of
Table~\ref{table:composit}: AA ($\Lambda_{\rm AA}^+$=55 TeV),
VV ($\Lambda_{\rm VV}^+$=55 TeV), LL ($\Lambda_{\rm LL}^+$=44 TeV),
RR ($\Lambda_{\rm RR}^+$=44 TeV), LR ($\Lambda_{\rm LR}^+$=21 TeV);
${\rm TeV}^{-1}$ model ($M_C$=12 TeV) and ADD$\pm$ models
($\Lambda_H$=4 TeV). Here, $\Lumint(e^-e^-)\approx\frac{1}{3}\Lumint(e^+e^-)$
has been assumed.}
\end{center}
\end{figure}
\par
Basically, a $\chi^2$ analysis of the differential cross section of
processes (\ref{bhabha})--(\ref{moller}) can be performed by dividing the
angular range into bins and introducing the sum over bins:
\begin{equation}
\chi^2({\cal O})= \sum_{\rm bins}\left(\frac{\Delta({\cal O})^{\rm
bin}} {\delta{\cal O}^{\rm bin}}\right)^2, \label{chi}
\end{equation}
where the relative deviations $\Delta({\cal O})$ are defined in
Eq.~(\ref{relat}) and $\delta{\cal O}$ denotes the expected experimental
relative uncertainties, that combine statistical and systematic ones.
\par
As a criterion to constrain the individual models, in particular to set the
discovery reach on the relevant mass scales, one essentially looks for
the smallest values of such parameters above which the deviation from the SM
prediction is too small to be observable within the experimental accuracy.
This value, indicated by non-observation of deviations, results from the
condition
\begin{equation}
\chi^2\le\chi^2_{\rm CL}, \label{chi2cl}
\end{equation}
where $\chi^2_{\rm CL}$ represents a critical value and we will take
$\chi^2_{\rm CL}=$3.84 for 95\% C.L.
\par
To make contact to the foreseeable experimental situation, we
impose cuts in the forward and backward directions of processes
(\ref{bhabha})--(\ref{moller}). Specifically, for Bhabha and
M{\o}ller scattering we consider the cut angular range
$\vert\cos\theta\vert< 0.9$ and divide it into nine equal-size
bins of width $\Delta\cos\theta =0.2$. Similarly, for the
annihilation into muon and tau pairs the angular range
$\vert\cos\theta\vert< 0.98$ will be considered. To assess the
statistical uncertainty, we assume the reconstruction efficiency
$\epsilon=95\%$ for final $l^+l^-$ events ($l=\mu,\,\tau$), and
$\epsilon\simeq 100\%$ for final $e^+e^-$ and $e^-e^-$ pairs.
Also, to assess the dependence of discovery reaches on the c.m.
energy and on the time-integrated luminosity, in the sequel an
ILC with $\sqrt s=0.5$ TeV and 1 TeV will be considered, with
${\cal L}_{\rm int}(e^+e^-)$ ranging from 100 ${\rm fb}^{-1}$ up
to 1000 ${\rm fb}^{-1}$.
\par
Concerning systematic uncertainties, an important source is
represented by the uncertainty on beams polarizations, for which
we assume $\delta P^\pm/P^\pm=0.2\%$ (and $\delta
P^-_{1,2}/P^-_{1,2}=0.2\%$ for the case of M{\o}ller scattering).
Also, a systematic error of 0.5\% in the luminosity determination
is assumed. In the case of the processes (\ref{bhabha}) and
(\ref{moller}), we analyze the following combinations of beams
polarizations: $(P^-,P^+)=(\vert P^-\vert,-\vert P^+\vert)$;
$(-\vert P^-\vert,\vert P^+\vert$; $(\vert P^-\vert,\vert
P^+\vert)$; $(-\vert P^-\vert,-\vert P^+\vert)$ with the
``standard'' envisaged values $\vert P^-\vert=0.8$ and $\vert
P^+\vert=0.6$ ($\vert P^-_1\vert=\vert P^-_2\vert=0.8$ for
M{\o}ller scattering). For the annihilation process (\ref{annil}),
we limit to the $(P^-,P^+)=(\vert P^-\vert,-\vert P^+\vert)$ and
$(-\vert P^-\vert,\vert P^+\vert)$ configurations. As for the
time-integrated luminosity, for simplicity we assume it to be
equally distributed between the different polarization
configurations defined above, and
$\Lumint(e^-e^-)\approx\frac{1}{3}\Lumint(e^+e^-)$ to account for
the reduction in luminosity of the $e^-e^-$ mode due to
anti-pinching in the interaction region \cite{Spencer:1996jf}. The
discovery limits are derived by taking the sum of the $\chi^2$ relevant to
the individual configurations of polarizations mentioned above and
imposing the constraint (\ref{chi2cl}). Also, we take into account
correlations between the different polarized cross sections, but
not those among individual angular bins.
\par
Regarding theoretical inputs, for the SM amplitudes we use
the effective Born approximation \cite{Consoli:1989pc} taking into
account electroweak corrections to propagators and vertices, with
$m_{\rm top}=175~\text{GeV}$ and $m_{\rm H} = 120~\text{GeV}$. Among the
${\cal O}(\alpha)$ QED corrections, the numerically most important
ones come from initial-state radiation. In the case of processes
(\ref{bhabha}) and (\ref{annil}), we account for that effect by a
using a structure function approach including both hard and soft
photon emission \cite{nicrosini} and the flux factor method
\cite{physicsatlep2}, respectively. To minimize the effect of
radiative flux return to the $s$-channel $Z$-exchange, a cut is applied
on the radiated photon energy, $\Delta\equiv E_\gamma/E_{\rm beam}<
1-M_Z^2/s$ with $\Delta= 0.9$, in order that interactions occur close
to the nominal collider energy and thus the best sensitivity to
the manifestations of non-standard physics can be obtained. Other QED
effects, such as final-state and initial-final state emission,
are found in process (\ref{annil}) to be numerically unimportant
for the chosen kinematical cuts, by using the ZFITTER
code \cite{Bardin:2001yd}. Regarding process (\ref{moller}), the
lowest-order corrections to the polarized cross sections are evaluated
by means of the FORTRAN code MOLLERAD \citer{Shumeiko1,Ilyichev:2005rx},
adapted to the present discussion.
\par
The numerical results for the discovery reaches on the effective
contact interaction mass scales, obtained by the above procedure in the case
of an ILC with $\sqrt s=0.5\hskip 3pt {\rm TeV}$ and
$\Lumint(e^+e^-)=100\hskip 3pt{\rm fb}^{-1}$, are summarized for the
different processes and polarization configurations in
Table~\ref{table:discov}. In this table, only the results for
positive interference between SM and non-standard interactions are
reported (i.e.,$\eta_{\alpha\beta}=+1$ for CI models and $\lambda=1$ for
ADD), because the sensitivity reach for negative interference is
practically the same. The results displayed in
Table~\ref{table:discov} relevant to the processes (\ref{bhabha})
and (\ref{annil}) with unpolarized beams are consistent with the
recent estimates in Ref.~\cite{Bourilkov:2003kj}. Also, the limits
on the CI mass scales resulting from $l^+l^-$ final states are based on the
assumption of $\mu-\tau$ universality.
\begin{table}
\begin{center}
\caption{Discovery reach (in TeV) on the mass scale parameters
(95\% C.L.) from the lepton pair production processes at $\sqrt
s=0.5$ TeV. For the $e^+e^-$ mode the three entries refer to
$\Lumint(e^+e^-)=100\ fb^{-1}$ and the polarizations
configurations $(\vert P^-\vert,\vert P^+\vert$)=(0,0); (0.8,0);
(0.8,0.6). For the $e^-e^-$ mode the configurations are $(\vert
P_1^-\vert,\vert P_2^-\vert)$=(0,0); (0.8,0); (0.8,0.8) and
$\Lumint(e^-e^-)\approx\frac{1}{3}\Lumint(e^+e^-)$.}
\vspace{0.2cm}
\begin{tabular}{|lr||c|c|c|c|}
\hline
    &  & \multicolumn{4}{|c|}{process} \\ \cline{3-6}
\multicolumn{2}{|c||}{model}  & $e^+e^- \to e^+e^-$ & $e^-e^- \to
e^-e^-$ & $e^+e^- \to \mu^+ \mu^-$ & $e^+e^- \to l^+l^-$ \\
\hline \hline


ADD$\pm$  & ($\Lambda_H$) &  4.1; 4.2; 4.3 & 3.8; 4.0; 4.1 & 2.8; 2.8; 2.9 & 3.0; 3.0; 3.2 \\

VV    & ($\Lambda $)  & 76.2; 80.8; 86.4 & 64.0; 68.8; 71.5 & 75.5; 76.4; 83.7 & 89.7; 90.7; 99.4 \\

AA    & ($\Lambda $)  & 47.4; 49.1; 69.1 & 58.0; 62.0; 64.9 & 67.3; 68.2; 74.8 & 80.1; 81.1; 88.9 \\

LL    & ($\Lambda $)  & 37.3; 45.5; 52.5 & 43.9; 52.4; 55.2 & 45.0; 51.0; 57.5 & 53.4; 60.5; 68.3 \\

RR    & ($\Lambda $)  & 36.0; 44.7; 52.2 & 42.3; 52.3; 55.4 & 43.2; 50.6; 57.5 & 51.3; 60.0; 68.3 \\

LR    & ($\Lambda $)  & 59.3; 61.6; 69.1 & 20.1; 22.1; 31.5 & 40.6; 46.0; 52.6 & 48.5; 55.0; 62.8 \\

RL    & ($\Lambda $)  & $\Lambda_{RL}=\Lambda_{LR}$ & $\Lambda_{RL}=\Lambda_{LR}$ & 40.8; 46.7; 53.4 & 48.7; 55.6; 63.6 \\

TeV   & ($M_C$)       & 12.0; 12.8; 13.8 & 11.7; 12.5; 12.9 & 16.8; 17.1; 18.7 & 20.0; 20.3; 22.2 \\
\hline

\end{tabular}
\end{center}

\label{table:discov}
\end{table}

Some features of the numbers in Table~\ref{table:discov} are
noteworthy. Firstly, there is complementarity of the leptonic
processes under consideration, (\ref{bhabha}), (\ref{annil}) and
(\ref{moller}), in the search for CI and ADD scenarios. Secondly,
for the chosen values of the time-integrated luminosity, the
discovery reaches on $\Lambda_H$ of the Bhabha and M{\o}ller
scattering processes can be larger than $\simeq 8\sqrt{s}$. Also, Bhabha
scattering is the process leading to the best search reach for
$\Lambda_H$, due to the available higher statistics in the data
sample in comparison to either M{\o}ller scattering or, to a large
extent, to the muon pair production process.
\par
One may recall that, on the purely statistical basis, the
discovery reach on $\Lambda_H$ should be expected to scale
with $s$ and $\Lumint$ like $\sim (s^3\Lumint)^{1/8}$, and the discovery
reach on the CI mass scales $\Lambda$s like $\sim (s\Lumint)^{1/4}$,
according to the different dimensions of the operators (\ref{dim-8})
and (\ref{CI}).

\section{Distinction  among the New Physics models}\label{sec:IV}
Let us assume one of the models to be consistent with data and
call it ``true'' model, for example the ADD model (\ref{dim-8})
with some value of $\Lambda_H$. We want to assess the level at
which this ``true'' model is distinguishable from the other ones,
that can compete with it as sources of corrections to the SM and
we call ``tested'' models, for any values of the corresponding mass
scale parameters. For example, we may take as ``tested'' model any one
of the four-fermion effective contact interactions listed in
Table~\ref{table:composit}. To that purpose, we can introduce relative
deviations of the differential cross sections from the ADD
predictions in each angular bin, arising from the CI models, analogous
to Eq.~(\ref{relat}):
\begin{equation}
{\tilde\Delta} ({\cal O})=
\frac{{\cal O}({\rm CI})-{\cal O}({\rm ADD})}{{\cal O}({\rm ADD})}.
\label{deviat}
\end{equation}
Correspondingly, a ${\tilde\chi}^2$ function analogous to Eq.~(\ref{chi})
can be introduced, with ${\tilde\delta} ({\cal O})$ defined in the
same way as $\delta ({\cal O})$ but, in this case, the statistical
uncertainty is referred to the ADD model and therefore depends on the
particular value of $\Lambda_H$. Since there can be ``confusion'' regions of
$\Lambda_H$ and $\Lambda_{\alpha\beta}$ values where also some CI
model can be consistent to the ADD predictions, on the basis of
such ${\tilde\chi}^2$ we can study whether these ``tested'' models
can be excluded or not to a given confidence level, that we always
assume 95\%, once the ADD model has been assumed as ``true''.
Then, we scan all values of $\Lambda_H$ up to the discovery
reach.
\par
\begin{figure}[htb!]
\begin{center}
\epsfig{file=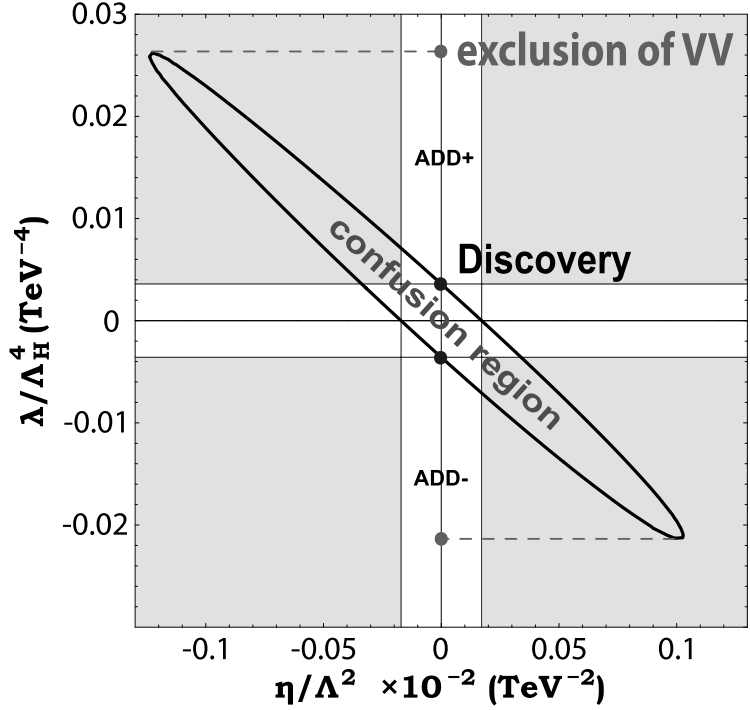,width=10.cm} 
\caption{Region of confusion  at 95\% C.L. for ADD and VV models
obtained from unpolarized Bhabha scattering at $\sqrt{s}=$0.5 TeV
and $\Lumint(e^+e^-)=$100 fb$^{-1}$.}
\end{center}
\label{confus}
\end{figure}

Thus, let us choose anyone of the ``tested'' CI models in
(\ref{deviat}), for definiteness the VV one defined in
Table~\ref{table:composit}, so that the ${\tilde\chi}^2$
mentioned above will be a function of the two parameters
$\epsilon_{\rm CI}\equiv\eta/\Lambda_{\alpha\beta}^2$ and
$\epsilon_{H}\equiv\lambda/\Lambda_H^4$ defined in Eqs.~(\ref{CI})
and (\ref{dim-8}), respectively. In Fig.~\ref{confus}, the four
gray areas in the ($\epsilon_{\rm CI}$, $\epsilon_{H}$) plane,
corresponding to the sign choices
($\eta,\lambda)=(1,1);\,(-1,-1);\,(1,-1);\,(-1,1)$, represent
values of the parameters for which both the ADD and the VV models
can give observable effects in unpolarized Bhabha scattering at
the ILC, with 95\% C.L. Also, the horizontal and vertical bands
correspond to the discovery reaches on the ADD and VV models at the
95\% C.L., derived in the previous section in the unpolarized case.
The ``confusion region'' is the area where the ${\tilde\chi}^2$ is
smaller than $\chi^2_{\rm CL}=$3.84 and the two models cannot be
distinguished at the 95\% C.L.
\par
As indicated in Fig.~\ref{confus}, one can find a maximal absolute
value of the scale parameter $\epsilon_{H}$ for which the ``tested'' VV
model hypothesis is expected to be excluded at the 95\% C.L. for any
value of the CI parameter $\epsilon_{\rm CI}$ taken in the gray area. We
denote the corresponding ADD mass scale parameter as $\Lambda_H^{\rm VV}$
and call it ``exclusion reach'' of the VV model.
\par
It is worth noticing that the ``confusion region'' is located only in the
areas with $(\eta,\lambda)=(1,-1)$ and $(-1,1)$. This is so because in these
regions the ADD$\pm$ and VV$\mp$ models provide the same sign of the leading
interference with the SM, hence of the deviation in the differential cross
sections as depicted in Fig.~1 and, therefore, they can mimic each other.
Conversely, there is no confusion region in the areas with
$(\eta,\lambda)=(1,1)$ and $(-1,-1)$ where interferences in the angular
distribution have opposite signs and therefore can easily be distinguished.
\par
Also, one may point out that, although this kind of $\chi^2$
analysis could in principle be applied to any observable, the
choice of the differential distribution as basic observable is
rather crucial to identify the ADD scenario. Integrated
observables such as, for example, the total cross section, the
forward-backward asymmetry and the left-right asymmetry, do not
allow to derive such a compact confusion region concentrated
around the origin as in Fig.~\ref{confus}. For instance, the confusion
regions determined from the integrated observables
would extend up to a few units of $\lambda/\Lambda_H^4$
(TeV$^{-4}$) and the corresponding reaches on $\Lambda_H$ would be
too much below the current discovery limits.
\begin{figure}[htb!]
\begin{center}
\epsfig{file=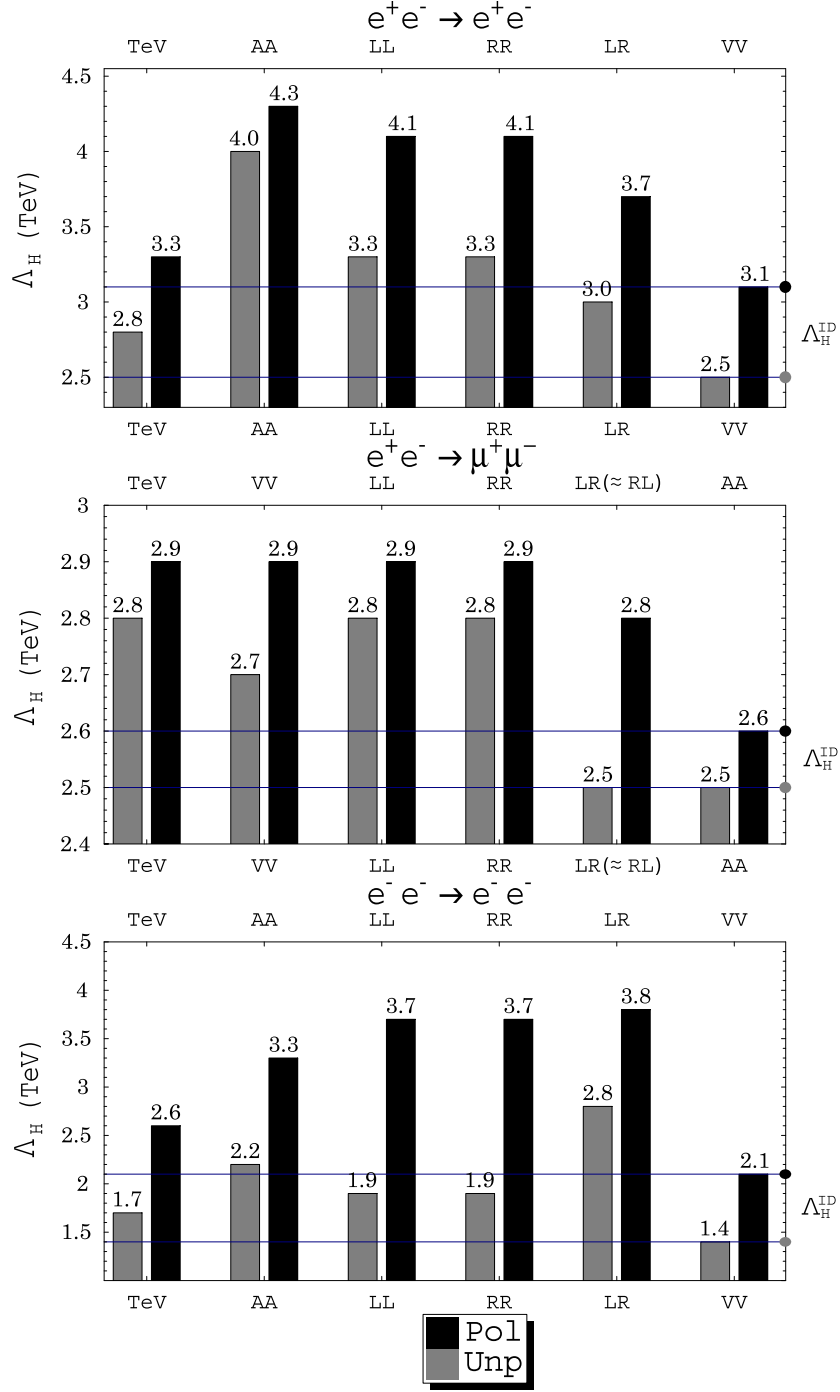,width=11.cm,height=18.5cm}
\vspace{0.0cm} \caption{\label{histogr}Exclusion and
identification reaches on $\Lambda_H$ at 95\% C.L. obtained from
the leptonic processes at $\sqrt{s}=$0.5 TeV,
$\Lumint(e^+e^-)=$100 fb$^{-1}$ for unpolarized beams (gray
histograms) and $\Lumint(e^-e^-)\approx\frac{1}{3}\Lumint(e^+e^-)$
and both beams polarized (black histograms) are illustrated.}
\end{center}
\end{figure}
The procedure outlined above can be repeated for all other types of effective
contact interaction models in Table~\ref{table:composit} as well as the
${\rm TeV}^{-1}$ gravity model mentioned in Table~\ref{table:epsilon},
and consequently one can evaluate the corresponding ``exclusion reaches''
$\Lambda_H^{\rm AA}$, $\Lambda_H^{\rm RR}$, $\Lambda_H^{\rm LL}$,
$\Lambda_H^{\rm LR}$ and $\Lambda_H^{\rm TeV}$. The
results of this kind of analysis for the three processes of interest here
with unpolarized beams (gray histograms) as well as polarized beams (black
histograms), are represented in Fig.~\ref{histogr}. In this figure, we have
considered an ILC with $\sqrt s=0.5\hskip 3pt{\rm TeV}$, time-integrated
luminosity $\Lumint (e^+e^-)=100\hskip 3pt{\rm fb}^{-1}$ and, in the
polarized case, the values of the beams polarizations anticipated in the
previous section, i.e.: $\vert P^-\vert=0.8$ and $\vert P^+\vert=0.6$ (and
for process (\ref{moller}) one third of the $e^+e^-$ luminosity and
$\vert P_1^-\vert=\vert P_2^-\vert=0.8$).
\par
As the final step, the ``identification reach'' on the ADD
scenario can be defined as the minimum of the $\Lambda_H$
``exclusion reaches'', $\Lambda_H^{\rm ID}=min\{\Lambda_H^{\rm VV},\,
\Lambda_H^{\rm AA}, \Lambda_H^{\rm RR},\, \Lambda_H^{\rm LL},\,
\Lambda_H^{\rm LR},\, \Lambda_H^{\rm TeV}\}$ as indicated in
Fig.~\ref{histogr}. It is clear that taking
$\Lambda_H<\Lambda_H^{\rm ID}$ allows one to exclude {\it all}
composite-like CI models as well as the ${\rm TeV}^{-1}$-scale gravity
model. In the specific example of the ILC parameters assumed in
the derivation of the results of Fig.~\ref{histogr}, it turns out
that the ``identification reach'' on the ADD graviton exchange
model obtained from unpolarized (polarized) Bhabha scattering
corresponds to $\Lambda_H^{\rm ID}=$2.5 (3.1) TeV. The
identification reaches on $\Lambda_H$ obtained from the different leptonic
processes are summarized in Table~\ref{IDnominal}.

\begin{table}
\begin{center}
\caption{95\% C.L. identification reach on $\Lambda_H^{\rm ID}$ at
$\sqrt{s}=0.5$ TeV and $\Lumint(e^+e^-)=100$ fb$^{-1}$.}
\vspace{0.2cm}
\begin{tabular}{|c||c|c|}
\hline
          & $P^{-}=P^{+}=0$       & $|P^{-}|=0.8; |P^{+}|=0.6$   \\
  process & $\Lambda_H^{\rm ID}$ (TeV) & $\Lambda_H^{\rm ID}$ (TeV) \\ \hline \hline
  $e^+e^- \to e^+e^-$      & 2.5 & 3.1 \\ \hline
  $e^+e^- \to \mu^+ \mu^-$ & 2.5 & 2.6 \\ \hline
$e^+e^- \to l^+ l^-$ & 2.7 & 2.8 \\ \hline
 $\Lumint(e^-e^-)\approx\frac{1}{3}\,\Lumint(e^+e^-)$ & $P^{-}_{1,2}=0$ & $|P^{-}_{1,2}|=0.8$ \\ \cline{2-3}
  $e^-e^- \to e^-e^-$      & 1.4 & 2.1 \\ \hline
\end{tabular}
\end{center}
\label{IDnominal}
\end{table}
\par
\begin{figure}[htb!]
\leavevmode \centering
\includegraphics[width=10.cm, angle=0]{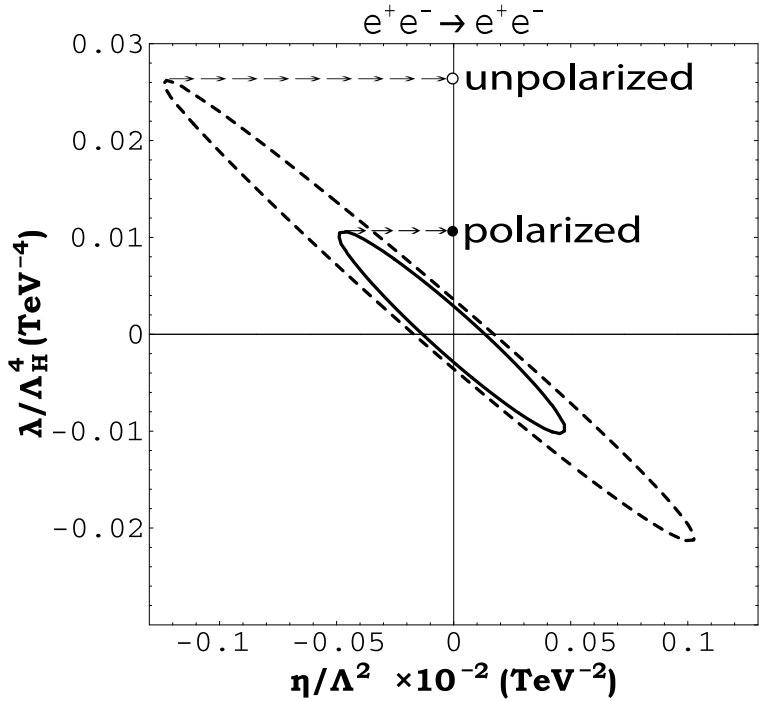}
\vspace*{0.cm}\caption{Region of confusion for ADD and VV models
obtained from unpolarized (dashed curve) and polarized (solid
curve) Bhabha scattering at $\sqrt{s}=$0.5 TeV and $\Lumint=$100
fb$^{-1}$.}
\label{IDpol}
\end{figure}

The r{\^o}le of beam polarization in increasing the sensitivity of the
leptonic processes to the graviton exchange effects, and particularly in
enhancing the potential of distinction from the other NP scenarios, is
essential, as exemplified in Fig.~\ref{IDpol} in the case of process
(\ref{bhabha}). Indeed, this figure indicates that the reduction of the
``confusion region'' with respect to the unpolarized case, allowed by
beams longitudinal polarization, can be quite substantial.
\par
\begin{figure}[htbp] 
\vspace*{-3.0cm} \centerline{ \hspace*{-0.6cm}
\includegraphics[width=10.0cm,angle=0]{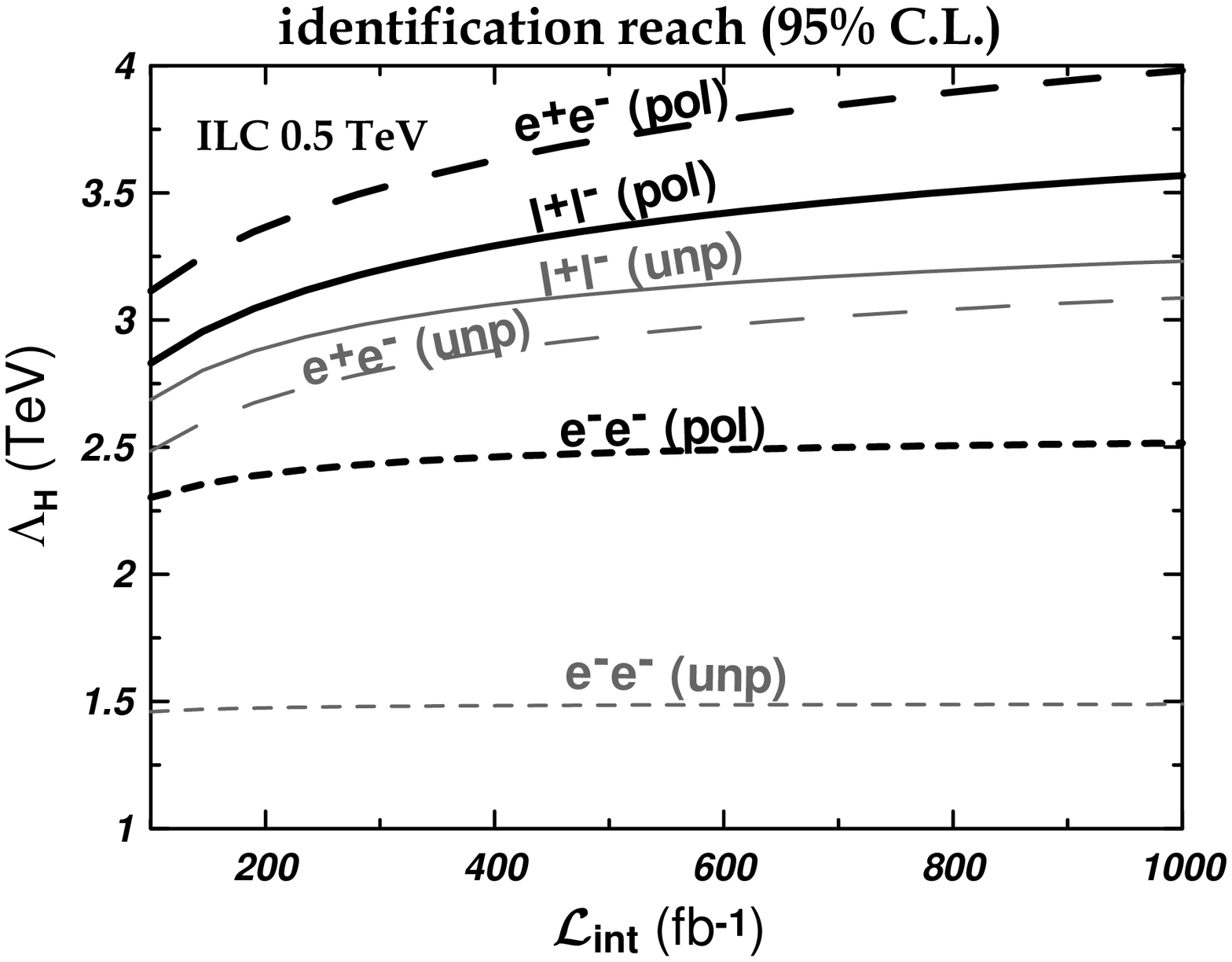}
\hspace*{-1.8cm}
\includegraphics[width=10.0cm,angle=0]{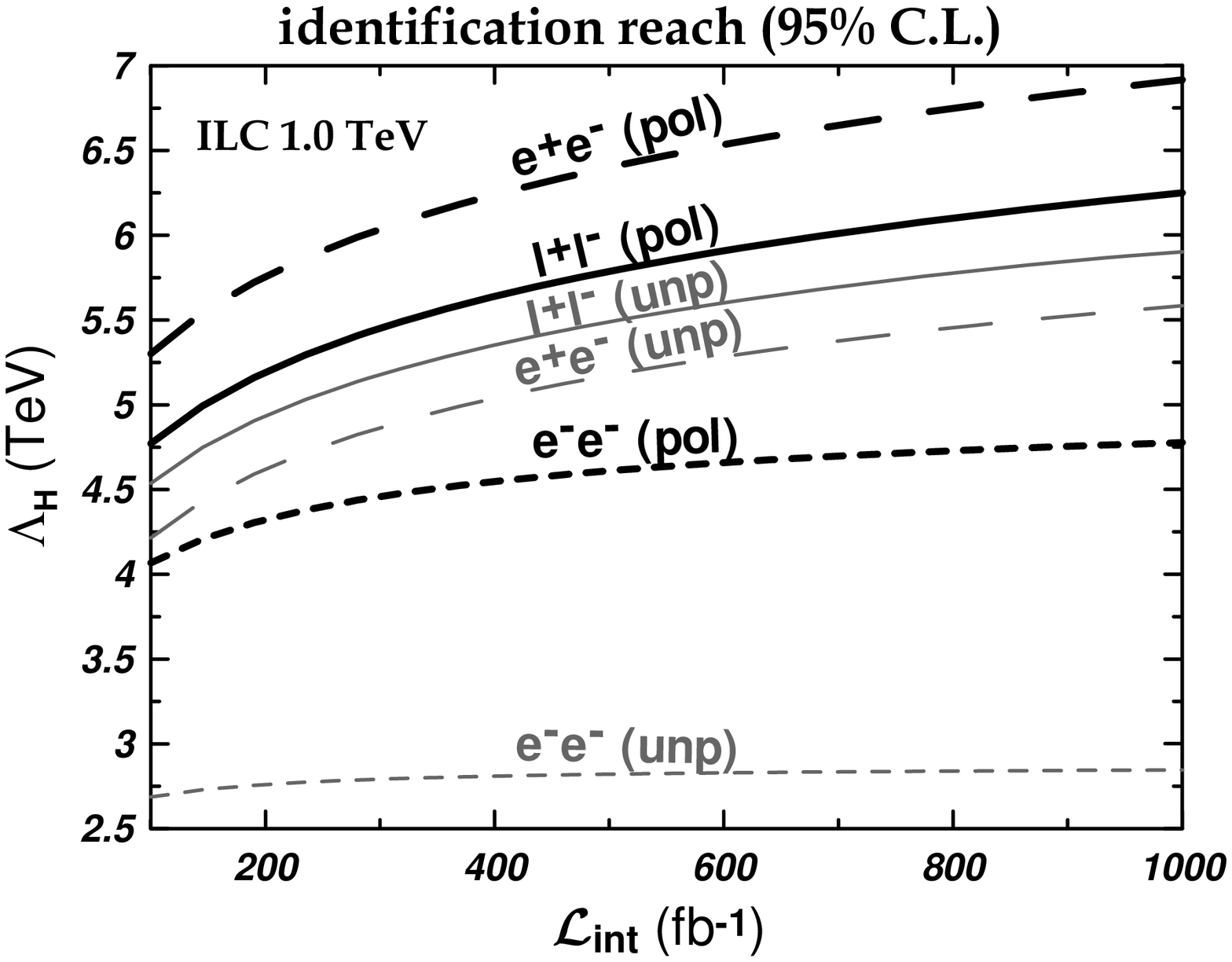}}
\vspace*{-4.0cm} \caption{95\% {\rm CL} identification reach on
the cutoff scale $\Lambda_H$ in the ADD model as a function of the
integrated luminosity obtained from lepton pair production
processes with unpolarized and both polarized beams at ILC(0.5
TeV) (left panel) and ILC(1 TeV) (right panel). }
\label{ID_ADD}
\end{figure}

Fig.~\ref{ID_ADD} shows the 95\% C.L. identification reach on the
graviton exchange mass scale $\Lambda_H$ as a function of
time-integrated luminosity\footnote{On the horizontal axis the
luminosity in the $e^+e^-$ channel is reported. One should recall
that 1/3 of that luminosity is assumed for the $e^-e^-$ mode.}
and two values of the c.m. energy,$\sqrt s=$ 0.5 TeV and 1 TeV.
The numerical procedure to obtain those results is the $\chi^2$
analysis described above, separately applied to the individual
processes (\ref{bhabha})--(\ref{moller}). In this figure, as well
as in the subsequent ones, curves are labelled by the final
states of the different processes, i.e., by $e^+e^-$ for Bhabha
scattering, $l^+l^-$ for the combination of $\mu^+\mu^-$ and
$\tau^+\tau^-$ production in the annihilation process, and
$e^-e^-$ for M{\o}ller scattering. The results for the
unpolarized differential cross sections are also shown for
comparison. In the case of polarization, the polarized cross
sections have been combined similar to the procedure followed in
the derivation of the discovery reaches, outlined in the previous
section and summarized in Table~\ref{table:discov}, with the
values of longitudinal polarizations exposed there.
\par
\begin{figure}[b]
\vspace*{-3.0cm} \centerline{ \hspace*{-0.6cm}
\includegraphics[width=10.0cm,angle=0]{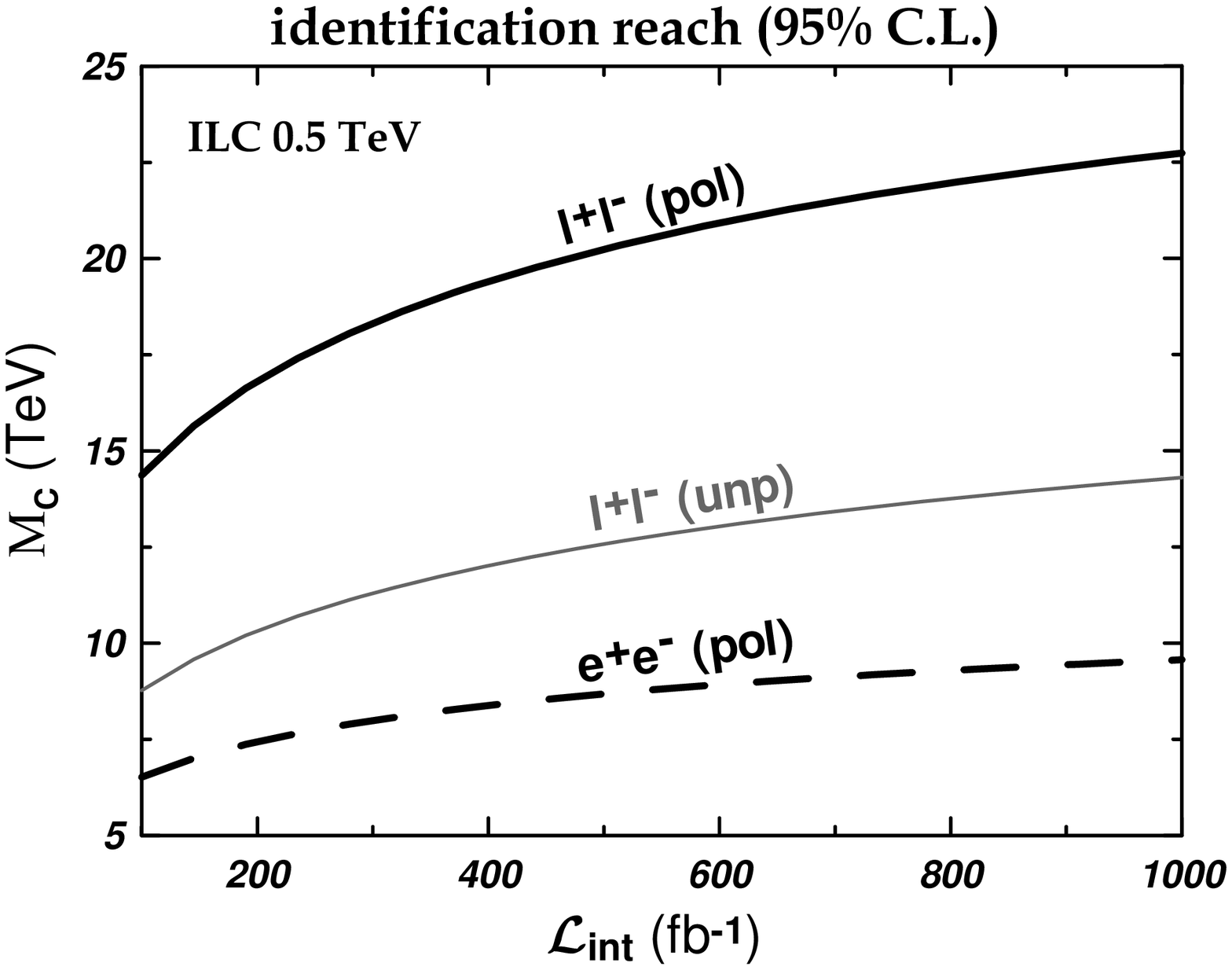}
\hspace*{-1.8cm}
\includegraphics[width=10.0cm,angle=0]{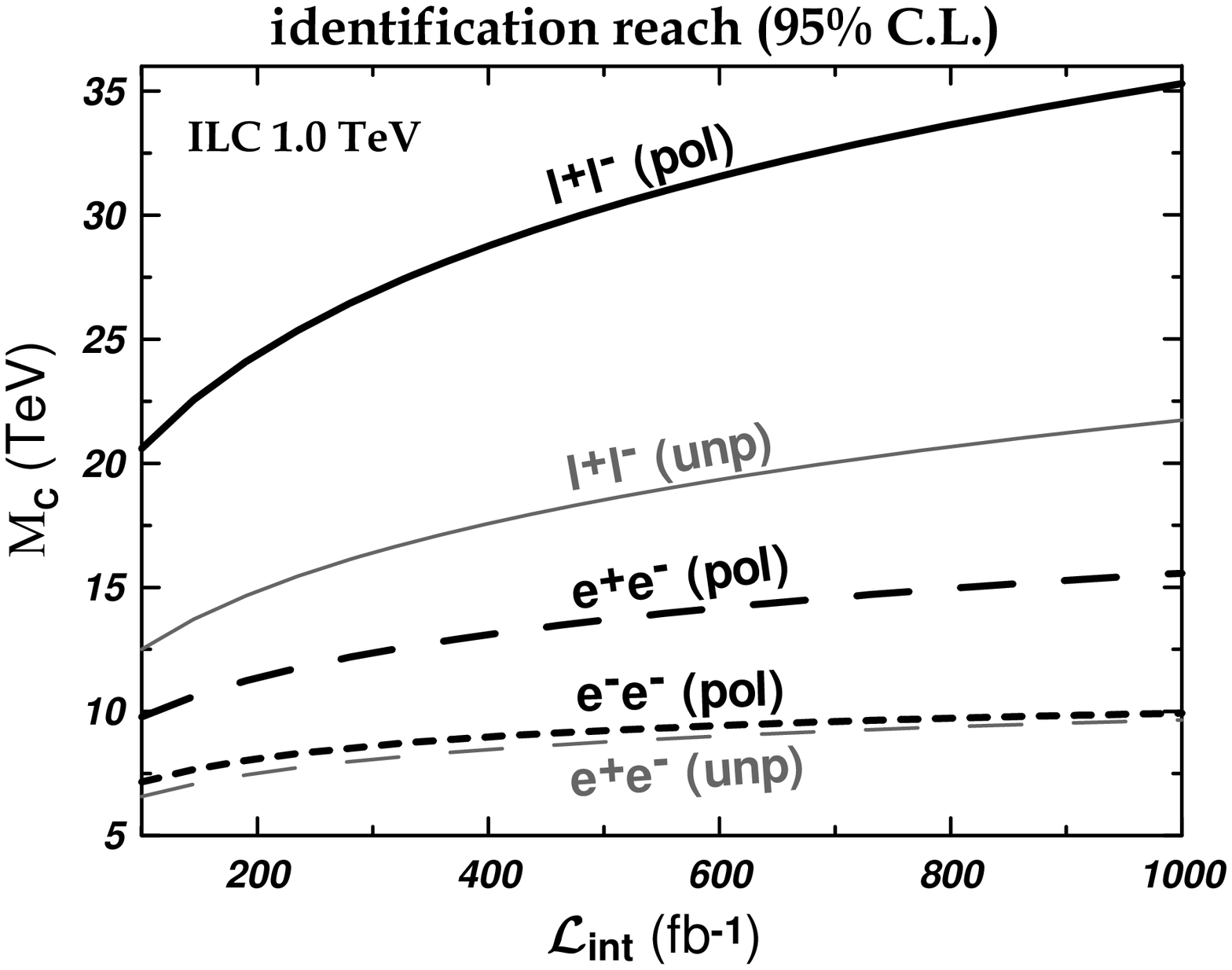}}
\vspace*{-4.0cm} \caption{Same as in Fig.{\ref{ID_ADD}} but for
the compactification scale $M_C$. }
 \label{ID_TeV}
\end{figure}
\begin{figure}[htbp]
\vspace*{-3.1cm} \centerline{ \hspace*{-0.6cm}
\includegraphics[width=10.0cm,angle=0]{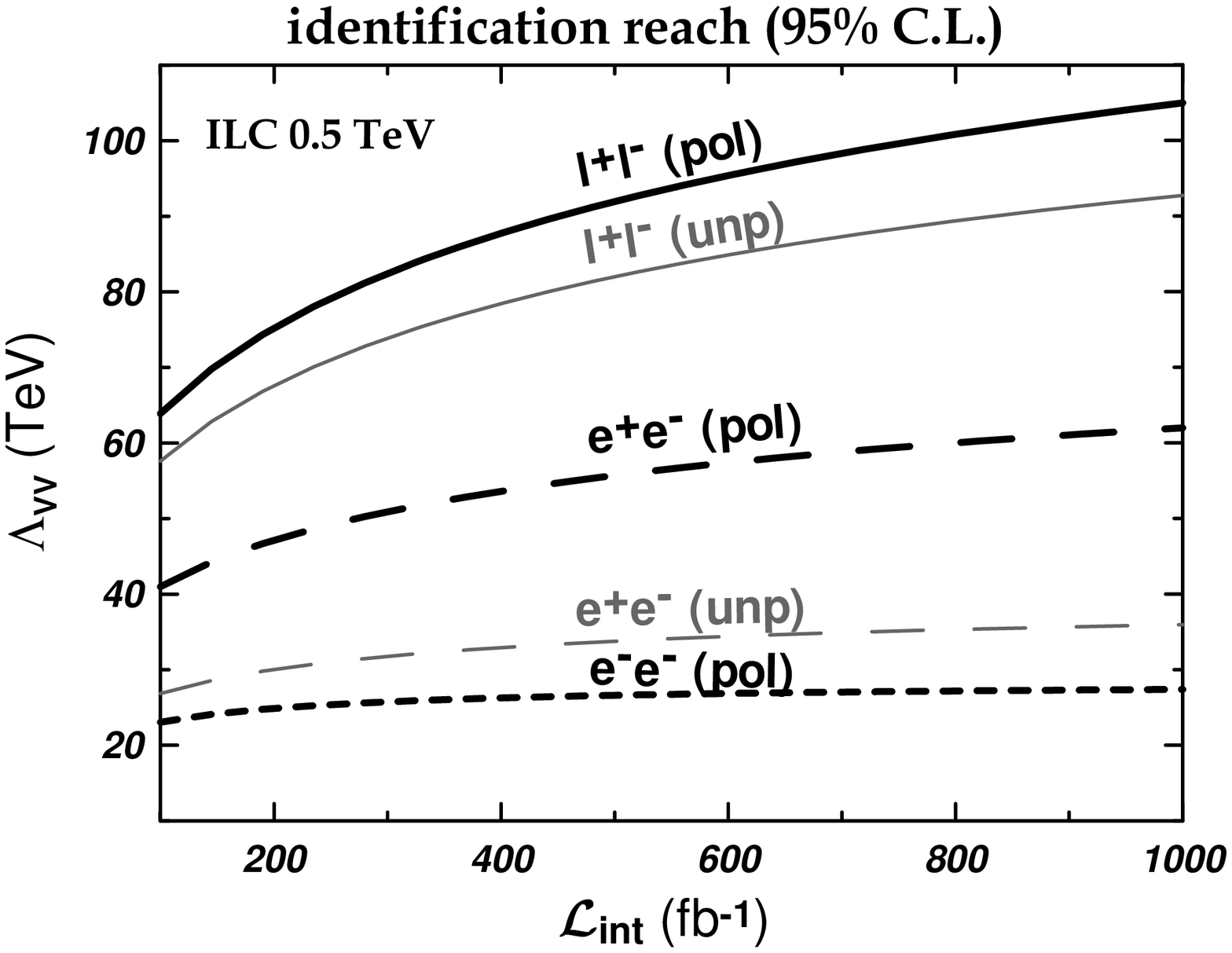}
\hspace*{-1.8cm}
\includegraphics[width=10.0cm,angle=0]{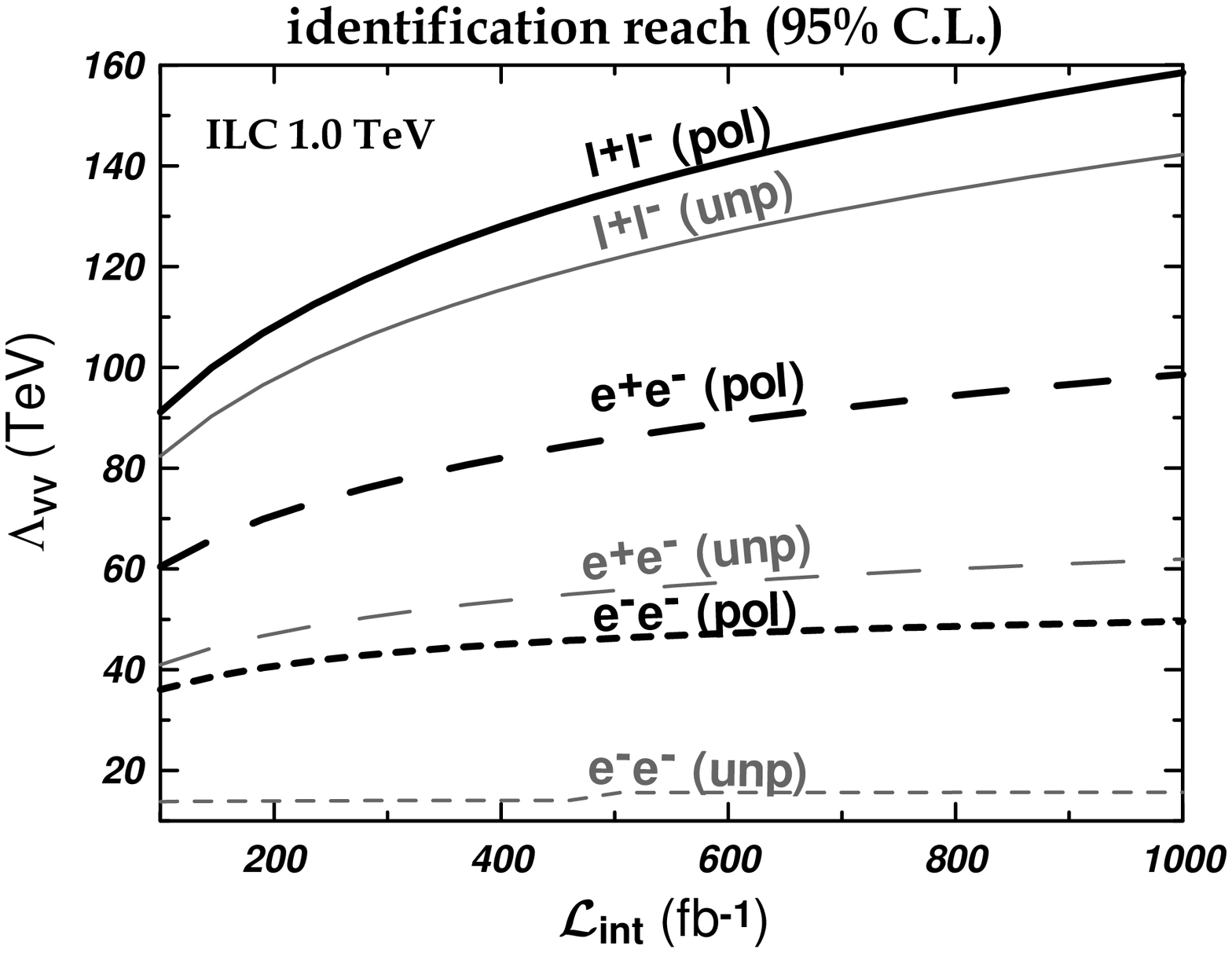}}
\vspace*{-7.0cm} \centerline{ \hspace*{-0.6cm}
\includegraphics[width=10.0cm,angle=0]{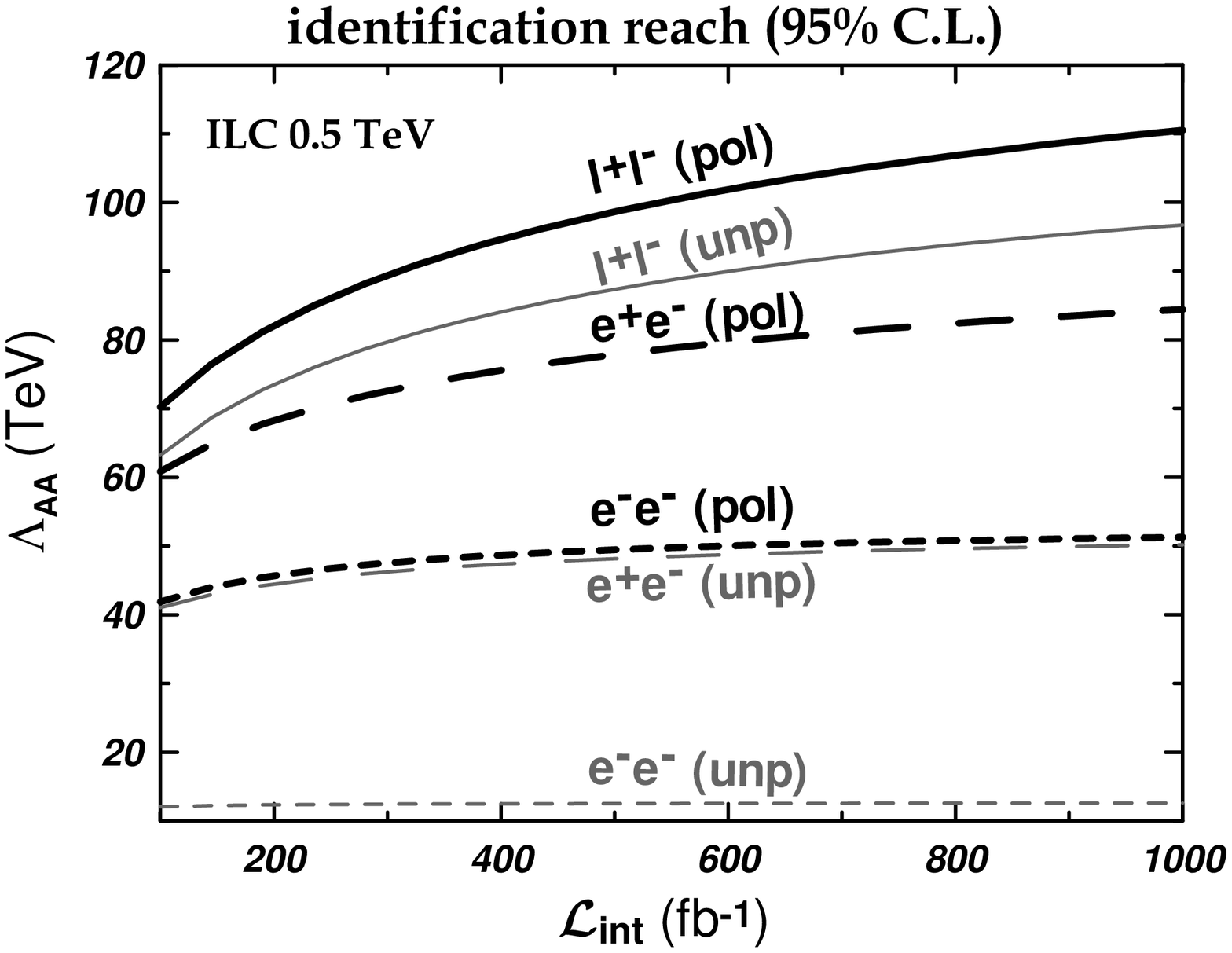}
\hspace*{-1.8cm}
\includegraphics[width=10.0cm,angle=0]{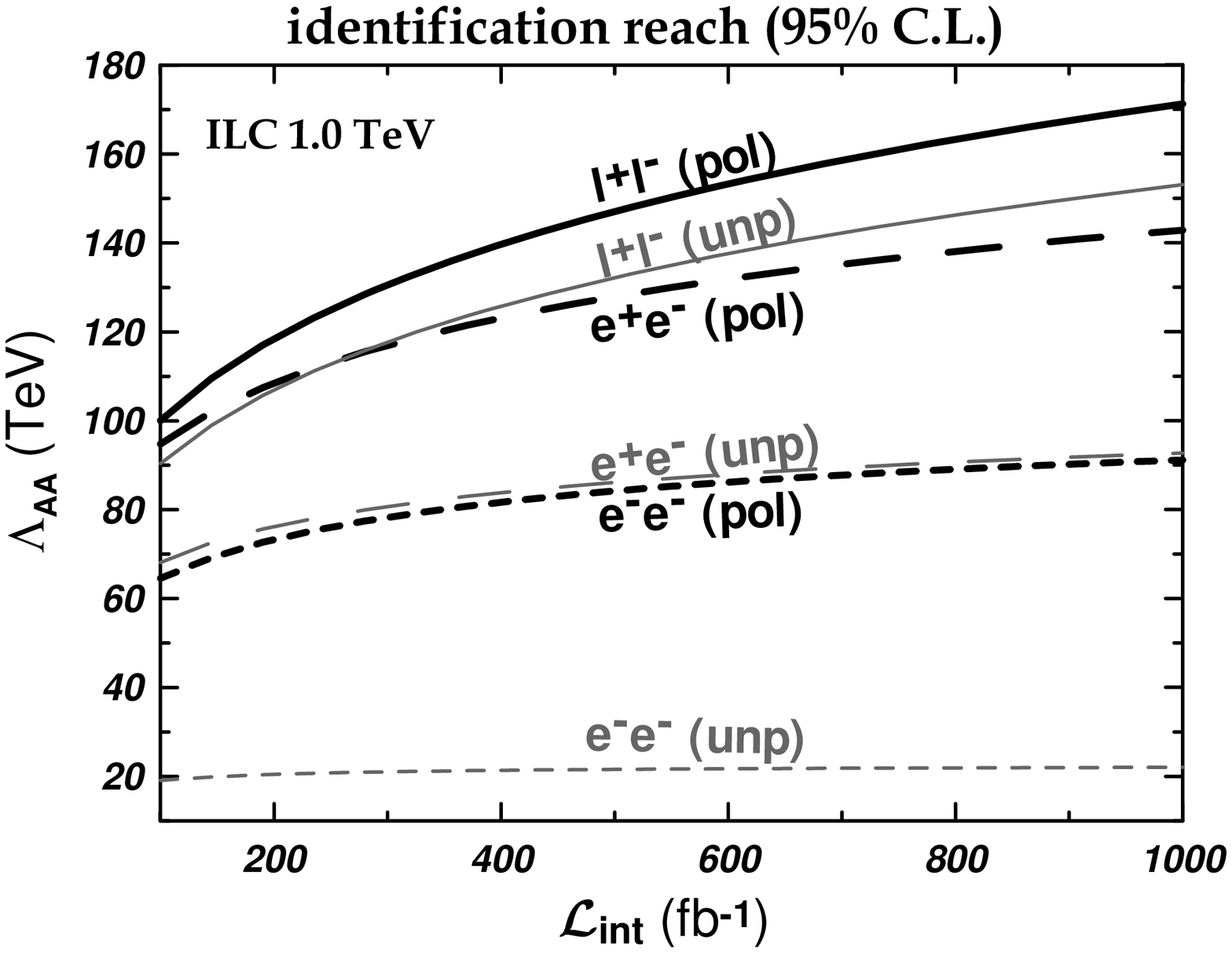}}
\vspace*{-7.0cm} \centerline{ \hspace*{-0.6cm}
\includegraphics[width=10.0cm,angle=0]{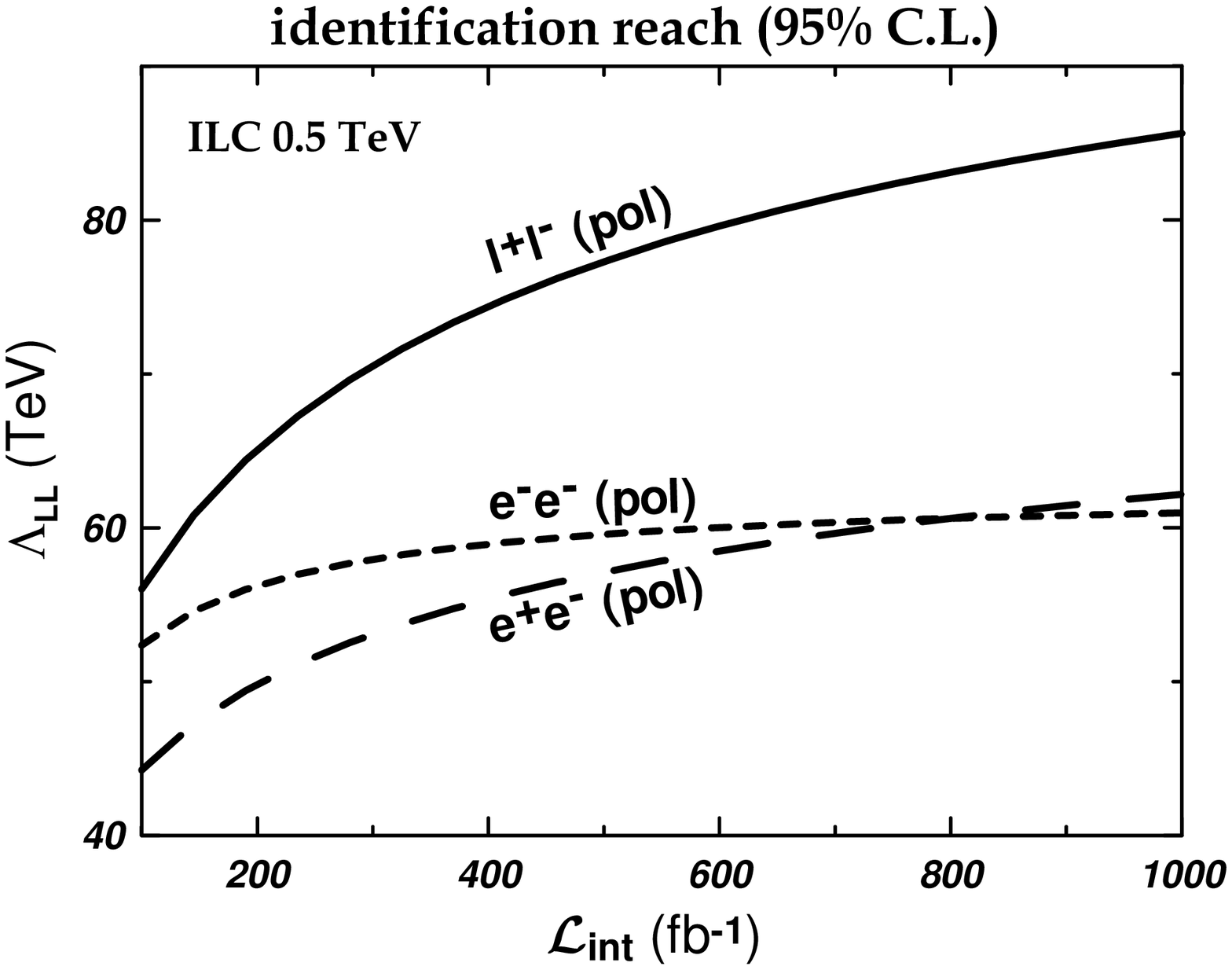}
\hspace*{-1.8cm}
\includegraphics[width=10.0cm,angle=0]{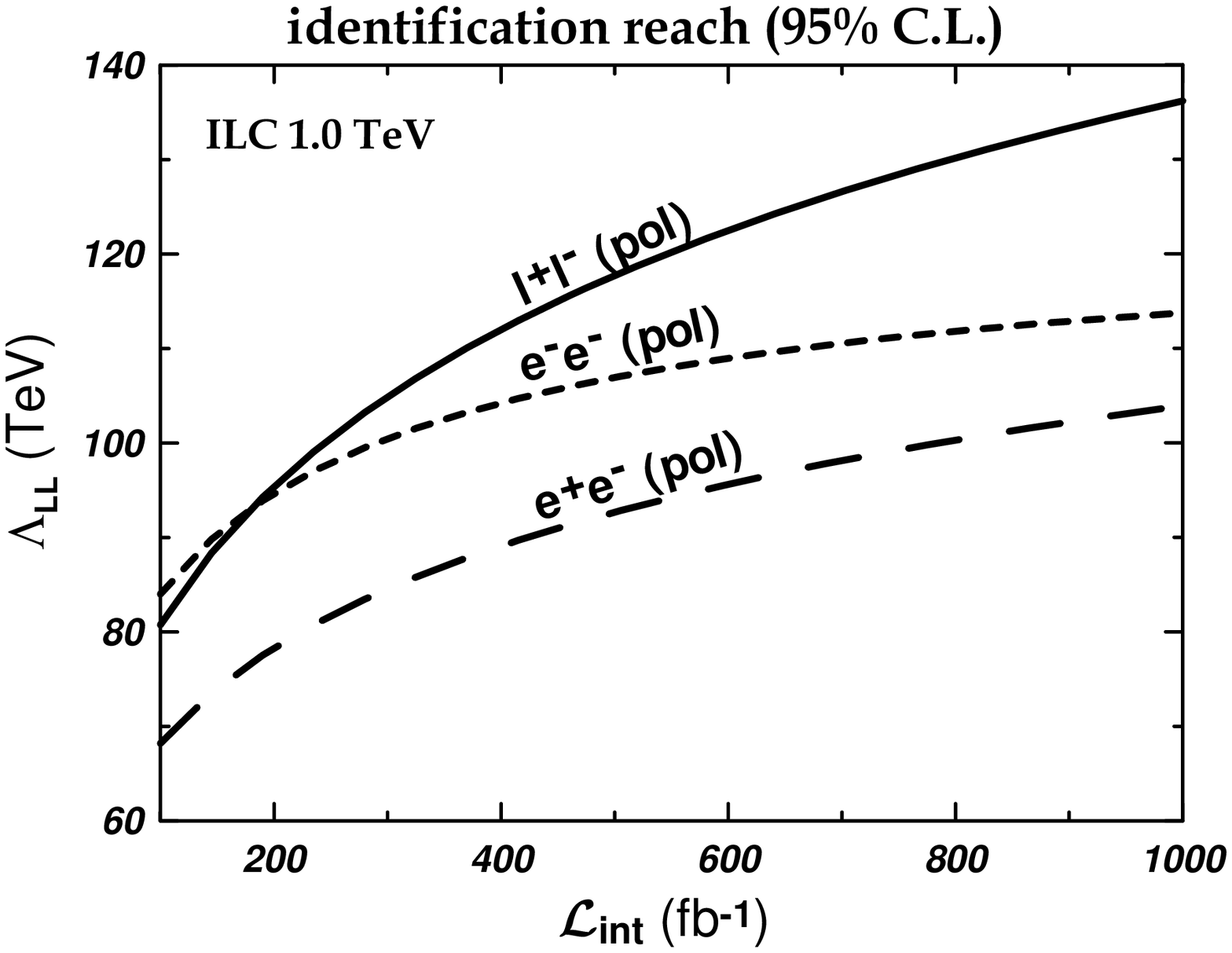}}
\vspace*{-4.0cm} \caption{Same as in Fig.{\ref{ID_ADD}} but for
the compositeness scale for $\rm VV$ (top panel), $\rm AA$
(central panel) and  $\rm LL$ (lower panel) type interactions. }
\label{ID_CI1}
\end{figure}

The identification reach on the ${\rm TeV}^{-1}$-scale gravity
model, obtained by applying the same kind of analysis to the three
lepton production processes of interest here, is presented in
Fig.~\ref{ID_TeV}, while the identification reaches on the
four-fermion contact interactions (\ref{CI}) are shown in
Figs.~\ref{ID_CI1} and \ref{ID_CI2}. In these figures, curves
corresponding to values of identification reaches that are found to fall
below the currently available experimental limits (listed in
sec.~\ref{sec:II}) have not been included.
\begin{figure}[htbp]
\vspace*{-3.1cm} \centerline{ \hspace*{-0.6cm}
\includegraphics[width=10.0cm,angle=0]{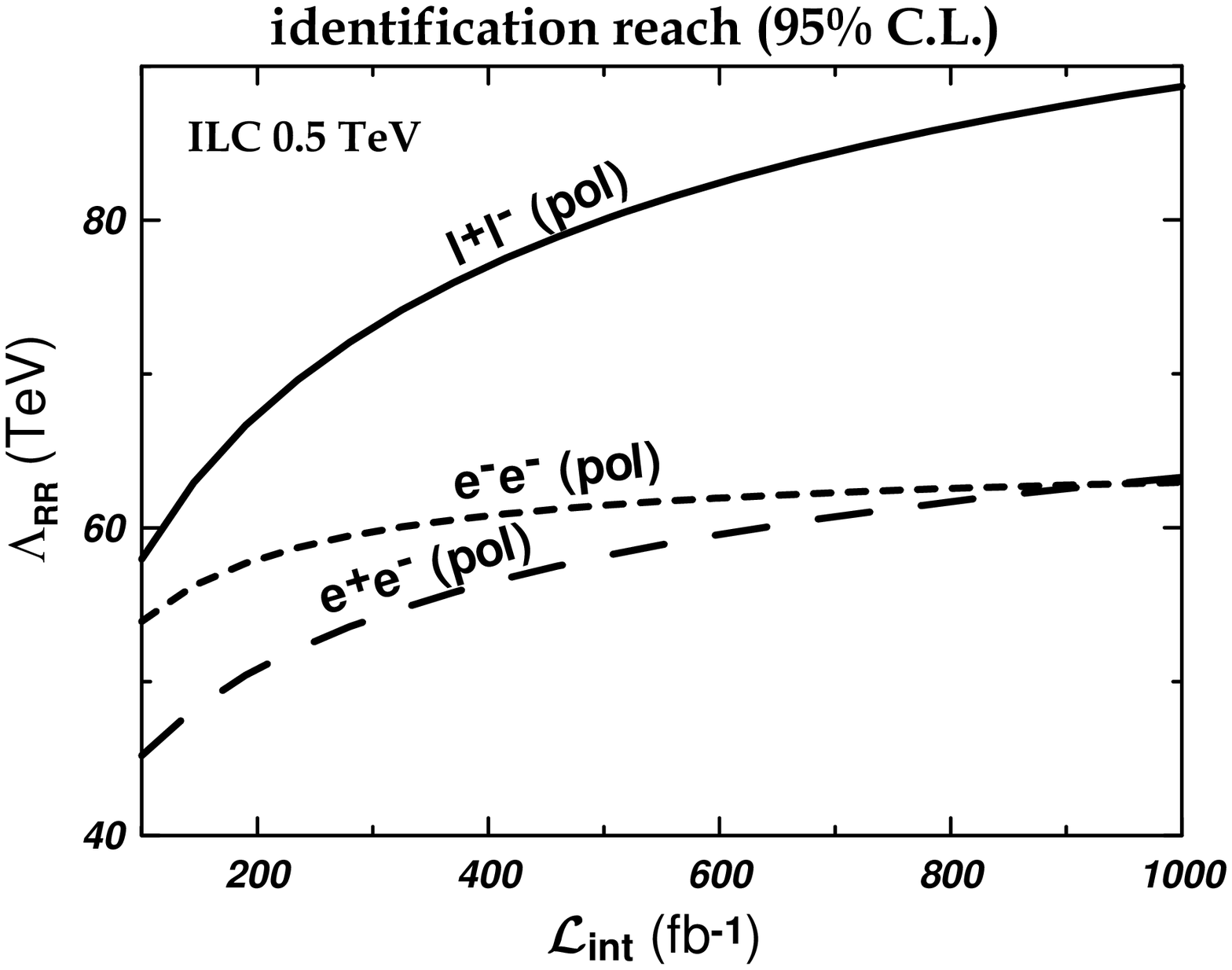}
\hspace*{-1.8cm}
\includegraphics[width=10.0cm,angle=0]{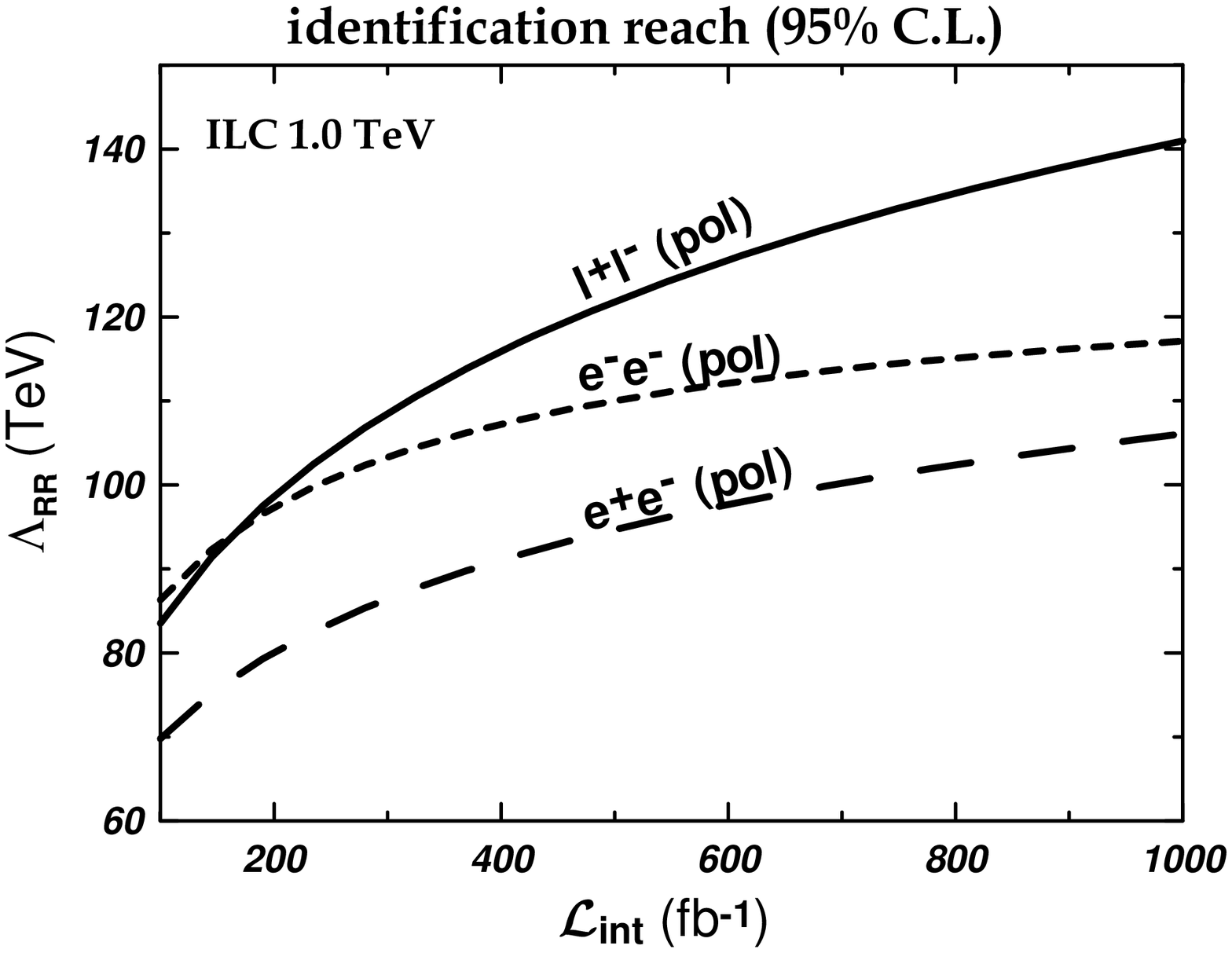}}
\vspace*{-7.0cm} \centerline{ \hspace*{-0.6cm}
\includegraphics[width=10.0cm,angle=0]{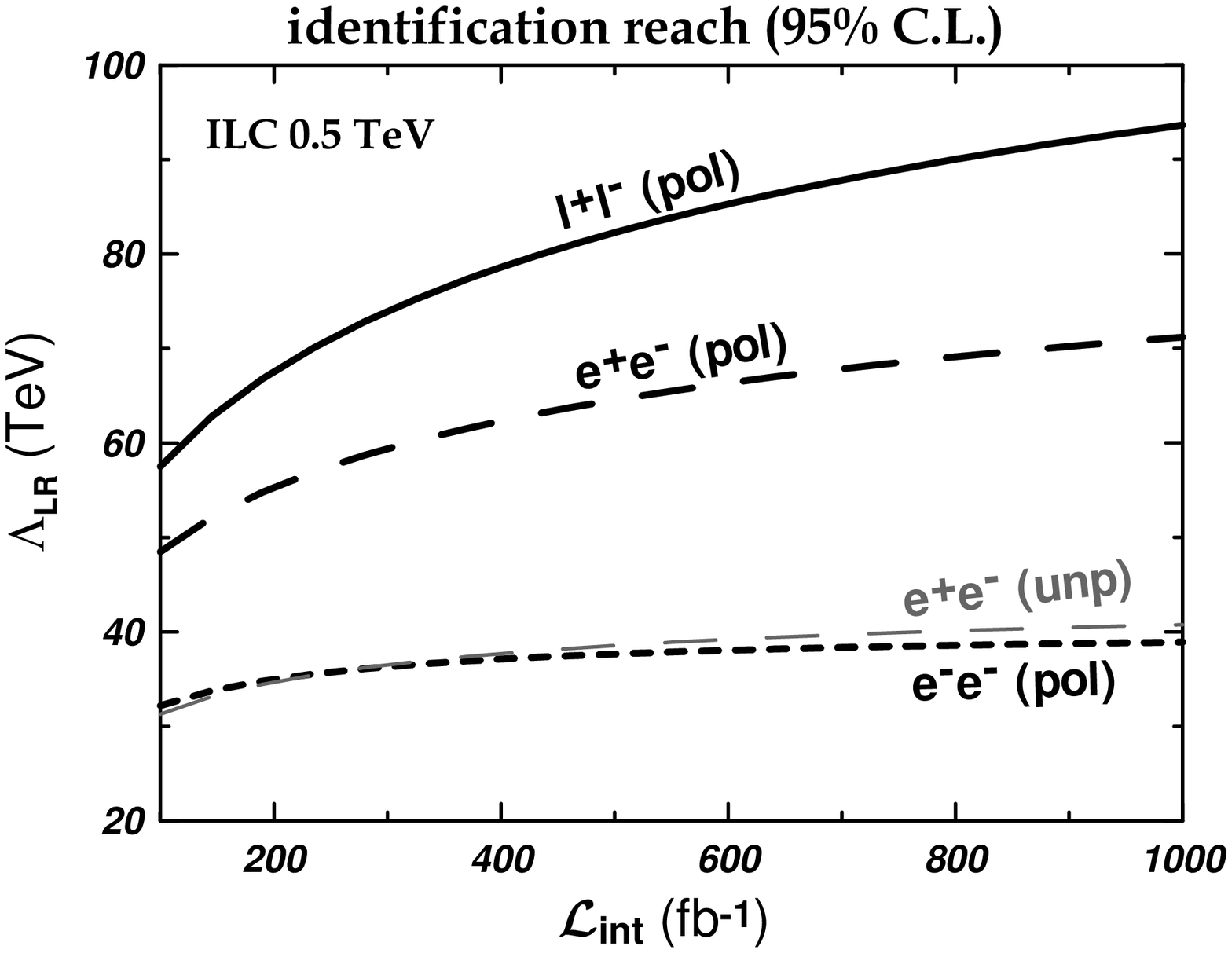}
\hspace*{-1.8cm}
\includegraphics[width=10.0cm,angle=0]{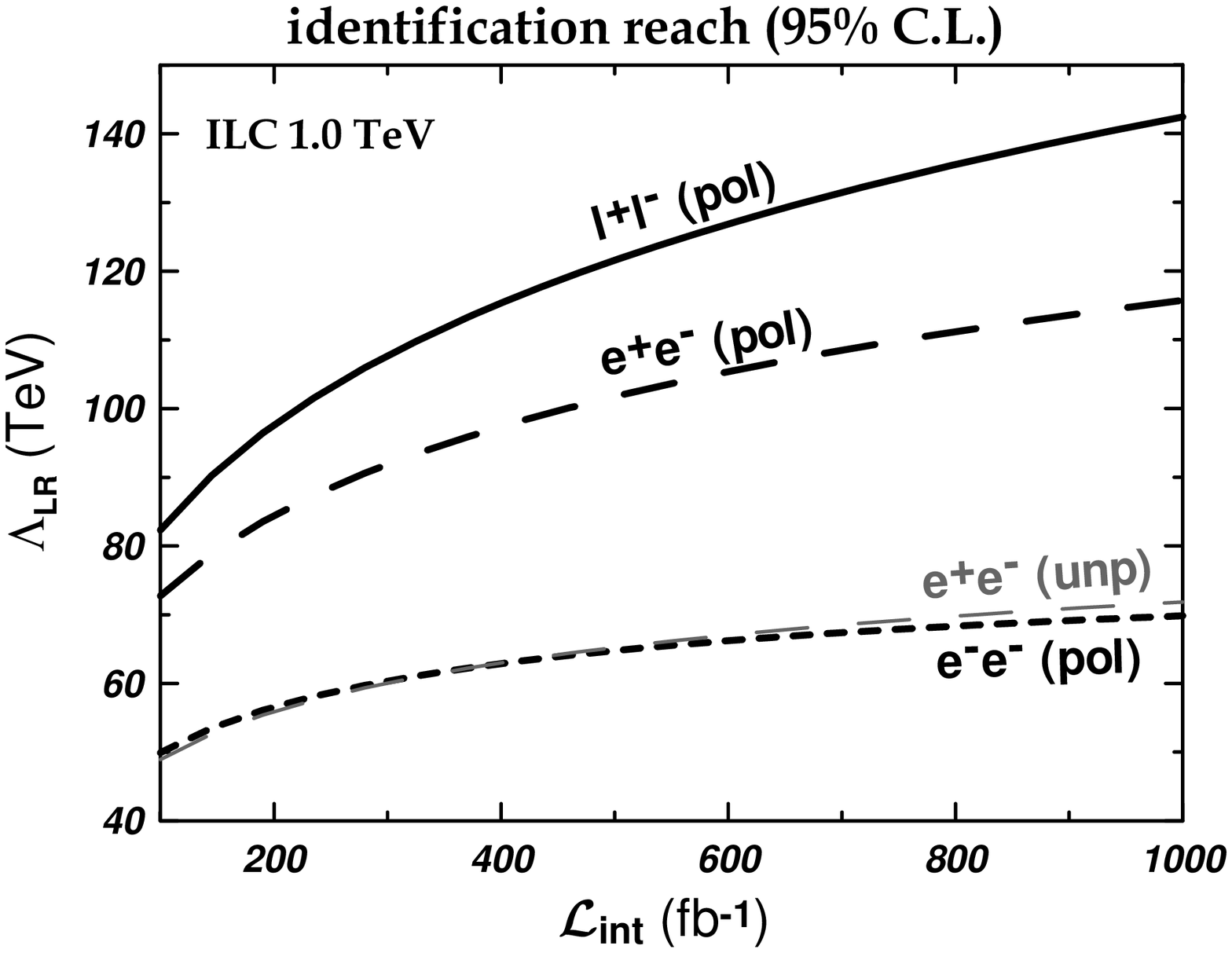}}
\vspace*{-7.0cm} \centerline{ \hspace*{-0.6cm}
\includegraphics[width=10.0cm,angle=0]{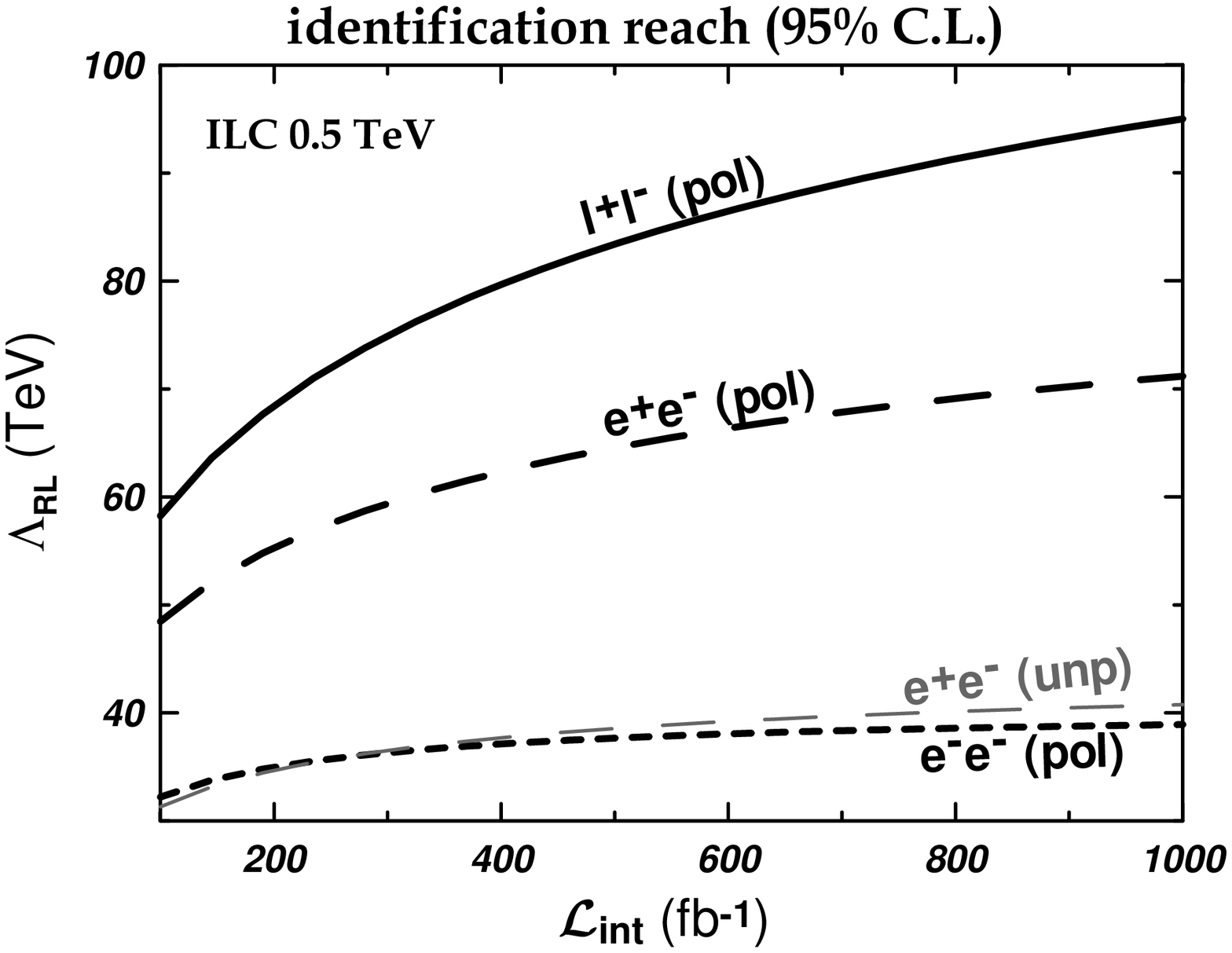}
\hspace*{-1.8cm}
\includegraphics[width=10.0cm,angle=0]{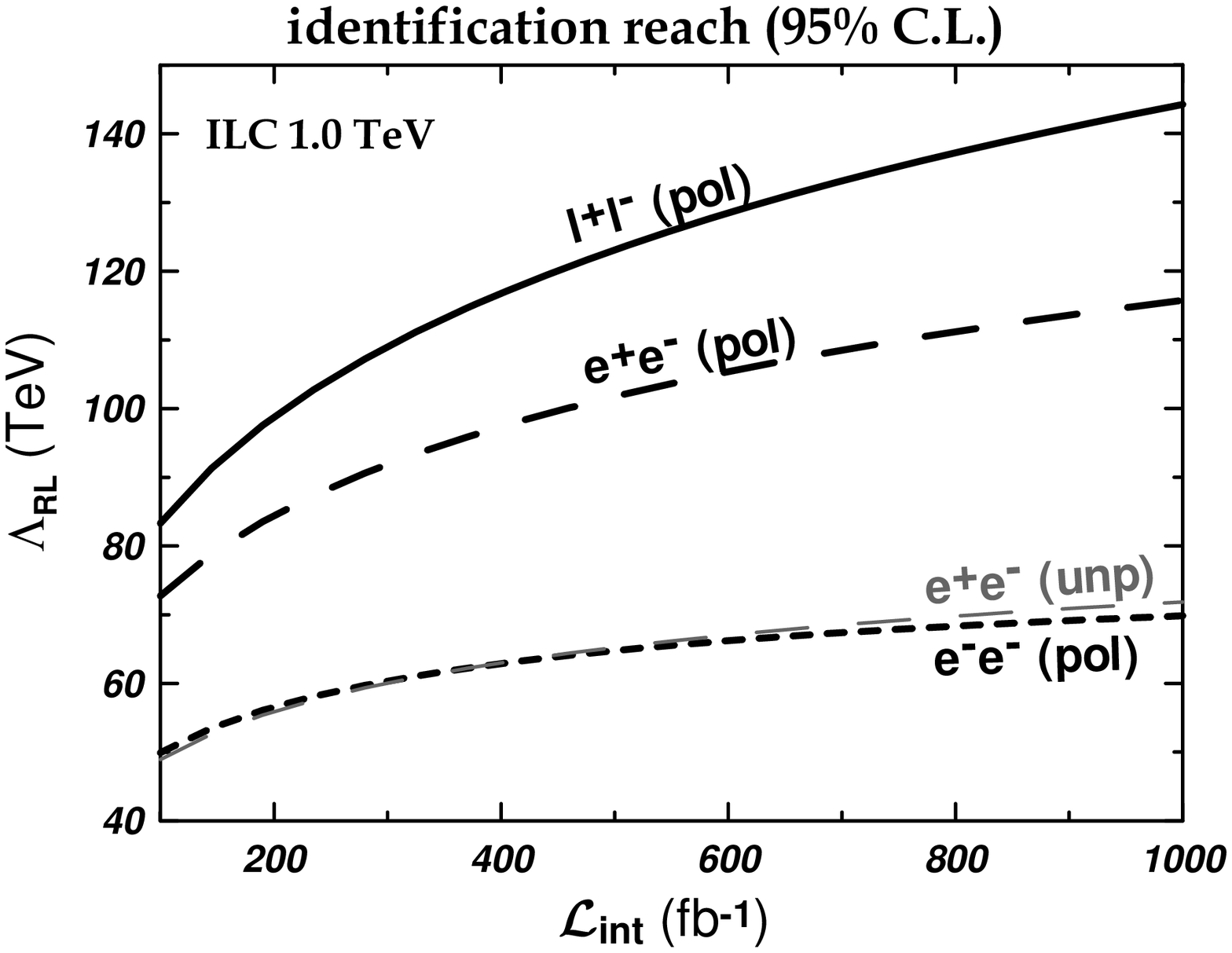}}
\vspace*{-4.0cm} \caption{Same as in Fig.{\ref{ID_ADD}} but for
the compositeness scale for $\rm RR$ (top panel), $\rm LR$
(central panel) and  $\rm RL$ (lower panel) type interactions. }
\label{ID_CI2}
\end{figure}
\section{Concluding remarks}\label{sec:V}
We here briefly comment on the main features of the findings in
Figs.\ref{ID_ADD}--\ref{ID_CI2} for the identification reaches on the
effective non-standard interactions considered in Sec.~\ref{sec:II},
obtainable from lepton pair production at the ILC with longitudinally
polarized beams.
\par
Fig.~\ref{ID_ADD} shows that, of the three considered processes,
Bhabha scattering definitely has the best identification
sensitivity on the scale $\Lambda_H$ characterizing the ADD model of
Eq.~{(\ref{dim-8})} for gravity in ``large'' compactified extra
dimensions. As one can see, in the polarized case, the
identification reach ranges from 3.1 TeV to 6.9 TeV, depending on
c.m energy and on luminosity. It could be of some interest in this
regard to estimate the ``resolving power'' on $\Lambda_H$
obtainable from the polarized process (\ref{bhabha}). This is
defined as the precision that in principle might be achieved on
the determination of $\Lambda_H$, in the case where the effects of
KK graviton exchange were observed. Fig.~\ref{RP} shows the
uncertainty obtainable on $\Lambda_H$ when we vary this parameter
in the range between the current experimental bound and the
expected identification reaches. Of course, the larger $\Lambda_H$
the worse the precision on it.
\begin{figure}[htb]
\refstepcounter{figure} \addtocounter{figure}{-1} \vspace{1cm}
\vspace{-2.0cm}
\begin{center}
\setlength{\unitlength}{1cm}
\begin{picture}(10.0,10.0)
\put(-3.0,-0.5)
{\mbox{\includegraphics[height=3.5in,width=3.1in]{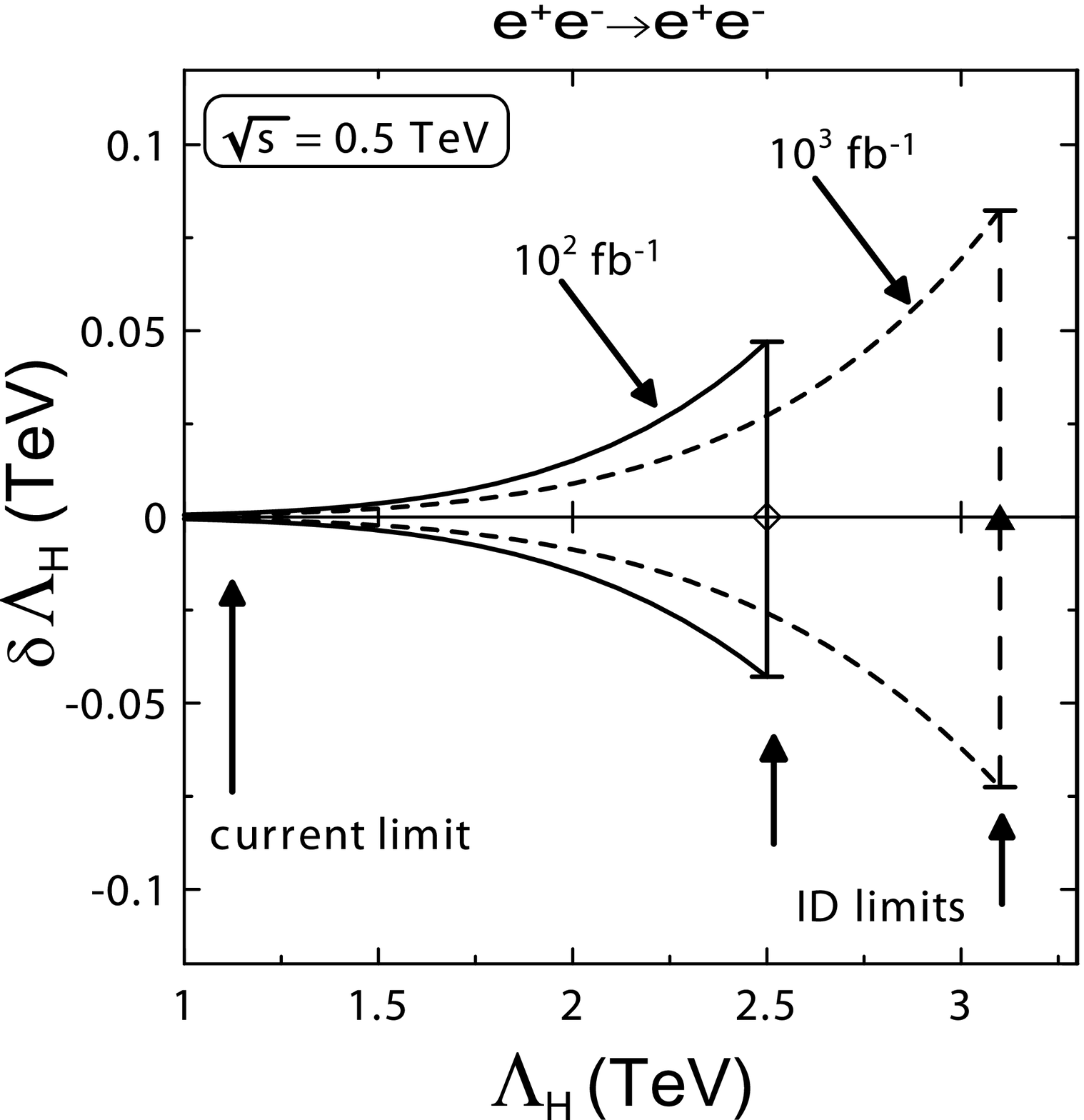}}
 \mbox{\includegraphics[height=3.5in,width=3.1in]{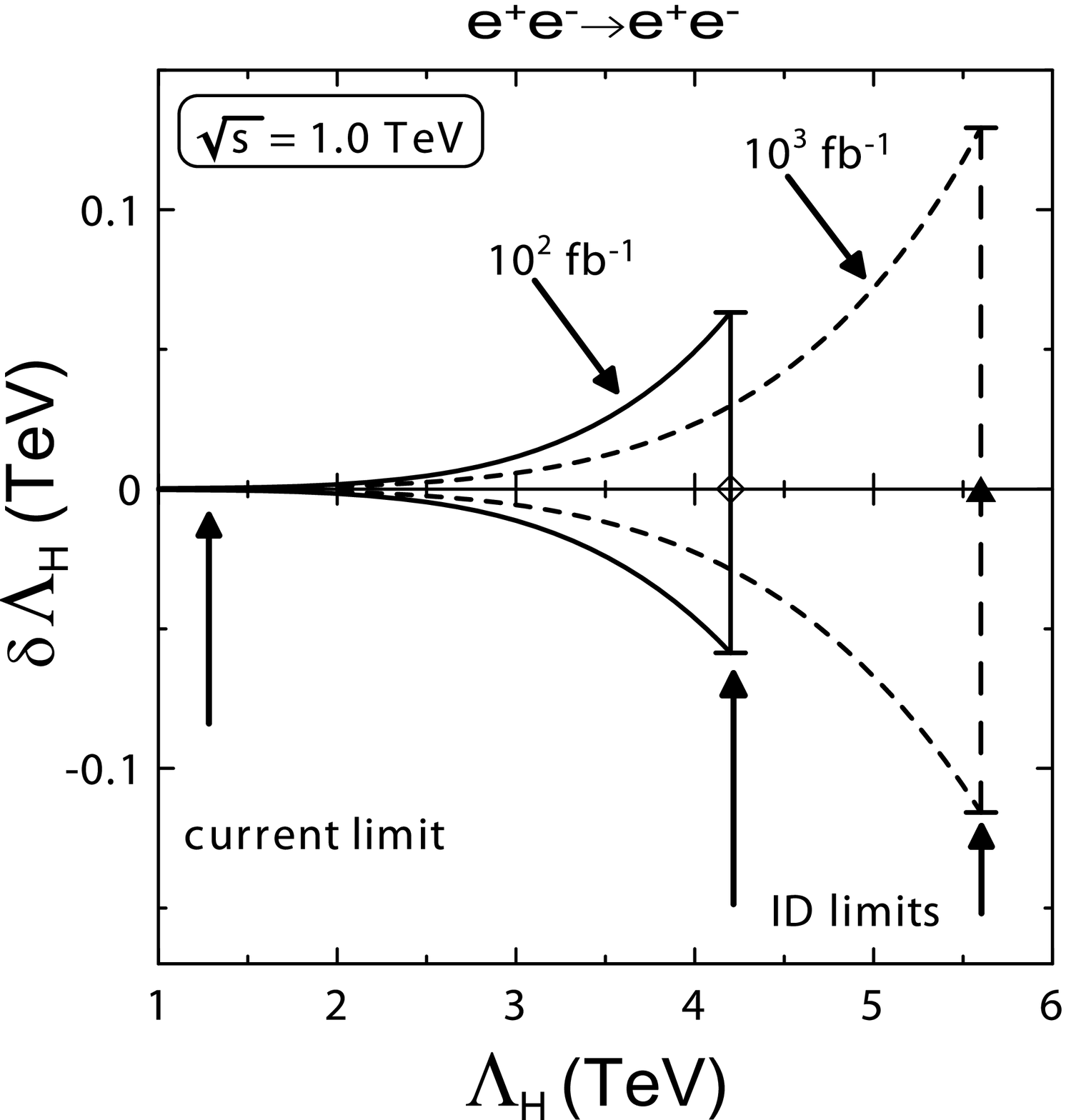}}}
\end{picture}
\vspace{.5cm} \caption{\label{FigPR} Resolving power $\delta
\Lambda_H$
 on the mass scales ($1\sigma$ level) {\it vs.}\ $\Lambda_H$
obtained from the polarized Bhabha scattering for the c.m. energy
$\sqrt{s}=$0.5 TeV (left panel) and 1 TeV (right panel) and
integrated luminosity of $\Lumint=100\hskip 2pt {\rm fb}^{-1}$ and
$1000\hskip 2pt {\rm fb}^{-1}$. } \label{RP}
\end{center}
\end{figure}
In contrast, although being competitive with the other processes for the
discovery reach, M{\o}ller scattering is found to give very poor
identification reaches on $\Lambda_H$
even in the polarized case, and appears to be saturated by the systematic
uncertainty as indicated by the corresponding curves quickly
becoming flat for increasing luminosity. Indeed, for this for process the KK
graviton exchange contribution to the RR and LL cross sections in
Eq.~(\ref{helsig}) is $\cos\theta$-independent therefore identical to
the CI interactions (\ref{CI}), while the $\cos\theta$-dependent contribution
to the LR cross section is considerably suppressed kinematically. Accordingly,
the potential of identification from the angular distributions is reduced.
This is qualitatively true in general, i.e., the identification reaches from
process (\ref{moller}) are rather moderate, and for some models well-below the
discovery reach.
\par
The $l^+l^-$ annihilation process (\ref{annil}) definitely provides the most
stringent identification reaches on ${\rm TeV}^{-1}$-scale gravity in extra
dimensions and on the four-fermion contact interactions (\ref{CI}) as well.
Quantitatively, Figs.~\ref{ID_TeV}--\ref{ID_CI2} show that the identification
reach on the compactification mass parameter $M_C$ ranges from 15 TeV to
35 TeV depending on the ILC energy and luminosity, and that the reaches on CI
four-fermion interactions (\ref{CI}) are rather high and significantly
enhanced by polarization: roughly, depending on the energy and the
luminosity, the identification reach on $\Lambda_{\rm VV}$ ranges
between 62 TeV and 160 TeV; that on
$\Lambda_{\rm AA}$ between 70 TeV and 170 TeV; for $\Lambda_{\rm LL}$
between 55 TeV and 135 TeV. Finally, for $\Lambda_{\rm RR}$,
$\Lambda_{\rm LR}$ and $\Lambda_{\rm RL}$ the discrimination range can be
between 57 TeV and 142 TeV.
\par
As regards the effective r{\^o}le of polarization in the
derivation of the identification reaches in
Figs.~\ref{ID_ADD}--\ref{ID_CI2}, in  particular that of the
positron beam polarization in addition to the electron one, it turns out
numerically that unpolarized beams do not lead to sensible ``identification
reaches'' on the mass scales relevant to the LL and RR four-fermion
CI models in the case of Bhabha and M{\o}ller scattering, and on the mass
scales of LL, RR, LR, RL models in the case of annihilation into $l^+l^-$.
This relates to the fact that $\vert g_{\rm L}^e\vert\simeq\vert g_{\rm
R}^e\vert$ so that the (leading) interferences of those NP
interactions with the SM have almost identical, hence in practice
indistinguishable, angular dependence (see Eq.~(\ref{helamp})).
Also, unpolarized M{\o}ller scattering may give identification
limits at the level of current experimental limits, for selected
values of energy and luminosity. In general then, for a given
process, polarized beams can have much higher
identification potential than the unpolarized ones. Specifically,
it turns out that the identification reaches in
Figs.~\ref{ID_ADD}--\ref{ID_CI2} for LL and RR models from Bhabha
scattering and those on LL, RR, LR, RL and gravity in ${\rm
TeV}^{-1}$ extra dimensions from annihilation into $l^+l^-$, are
mostly rely on electron polarization and not as much on the
simultaneous positron polarization. By contrast, the
identification reaches on ADD, AA, VV, LR and ${\rm TeV}^{-1}$
models from Bhabha scattering and those on ADD, AA and VV models
from annihilation into $l^+l^-$ pairs definitely need {\it both}
electron {\it and} positron polarization. It may also be noticed
that, in many cases, polarized cross sections allow to identify
the CI models almost near the discovery reaches (summarized for
the lowest considered values of energy and luminosity in
Table~\ref{table:discov}).

\vspace{0.5cm}
\leftline{\bf Acknowledgements}
\par\noindent
AAP acknowledges the support of INFN and of MIUR (Italian Ministry
of University and research). NP has been partially supported by
funds of MIUR and of the University of Trieste. This research has
been partially supported by the Abdus Salam ICTP and  the World
Federation of Scientists (National scholarship programme). AAP
and AVT also acknowledge the Belarusian Republican Foundation for
Fundamental Research.



\end{document}